\UseRawInputEncoding
\documentclass[a4paper,12pt]{article}
\pdfoutput=1 

\usepackage{jheppub} 
\usepackage{hyperref}
\usepackage{amsmath}
\usepackage{relsize}
\usepackage{psfrag}
\usepackage{epsfig}
\usepackage{xcolor}
\usepackage{color}
\usepackage{graphics}
\usepackage{url}
\usepackage{relsize}

\usepackage[T1]{fontenc} 
\usepackage[percent]{overpic}
 \normalsize
\newcommand{\bmat}{\left(\begin{array}}
\newcommand{\emat}{\end{array}\right)}
\newcommand{\be}{\begin{equation}}
\newcommand{\ee}{\end{equation}}
\newcommand{\bea}{\begin{eqnarray}}
\newcommand{\eea}{\end{eqnarray}}

\def\lsim{\raise0.3ex\hbox{$\;<$\kern-0.75em\raise-1.1ex\hbox{$\sim\;$}}}
\def\gsim{\raise0.3ex\hbox{$\;>$\kern-0.75em\raise-1.1ex\hbox{$\sim\;$}}}

\title{\boldmath No-scale gauge non-singlet inflation inducing TeV scale inverse seesaw 
mechanism}

\author[a]{Ahmad Moursy}

\affiliation[a]{Department of Basic Sciences, Faculty of Computers and Artificial Intelligence,  Cairo University, Giza 12613, Egypt.}

\emailAdd{a.moursy@fci-cu.edu.eg}

\abstract{ We develop a model of sneutrino inflation that is charged under $U(1)_{B-L}$ gauge symmetry, in no-scale supergravity framework. The model provides an interesting modification of tribrid inflation. We impose $U(1)_R$ symmetry on the renormalizable level while allow Planck suppressed non-renormalizable operators that break R-symmetry. This plays a crucial role in realizing a Starobinsly-like inflation scenario from one hand. On the other hand it plays an essential role, as well as SUSY breaking effects, in deriving the tiny neutrino masses via TeV inverse seesaw mechanism. Thus, we provide an interpretation for the extremely small value of the $\mu_S$ mass parameter required for inverse seesaw mechanism.  We discuss a reheating scenario and possible constraints on the model parameter space in connection to neutrino masses.  
}

\begin{document}
\maketitle
\flushbottom
\section{Introduction}
\label{sec:intro}
Remarkable studies over the past few decades were devoted to explore the connection and interplay between different hierarchical scales in nature. Finding such connections helps in understanding the fundamental origin of these scales. This issue has attracted careful studies on the huge hierarchy between the scales at the low energy observable sector, such as electroweak and neutrino mass scales from one hand, and on the other hand the early universe scales such as Planck, GUT, SUSY breaking and inflation scales.

Inflation paradigm is supported by CMB observations. Planck collaboration \cite{Akrami:2018odb} has confirmed the values of the spectral index of the scalar fluctuations $n_s=0.955-0.974$, up to 2 sigma exclusion limits, and the tensor to scalar ratio $r< 0.08$. One of the appealing models of inflation is the Starobinsky model \cite{Starobinsky:1980te} whose predictions are consistent with Planck observations. Starobinsky-like potentials can be implemented in supergravity framework such as no-scale supergravity models \cite{Cremmer:1983bf,Ellis:1984bm,Ellis:2013xoa,Ellis:2013nxa,Ellis:2017jcp,Khalil:2018iip,Moursy:2020sit}. The latter models have a striking feature of solving naturally the $\eta$-problem without any need to preserve R-symmetry that is required in supersymmetric hybrid inflation models \cite{Dvali:1994ms,Antusch:2004hd}.

It is tempting to study the link between inflation scale and neutrino mass scale. The detected neutrino oscillations \cite{Super-Kamiokande:1998kpq,LSND:2001aii,K2K:2002icj,T2K:2011ypd,DayaBay:2013yxg} implied that neutrinos have tiny masses. The latter hinted at new physics beyond the standard model that accounts for seesaw mechanisms \cite{Minkowski:1977sc,Mohapatra:1979ia,Yanagida:1979as,Schechter:1980gr}. Interestingly right-handed neutrino supermultiplets can be employed to play a double role in realizing seesaw type I mechanism using its fermionic components, while the scalar component (right-handed sneutrino) can be assigned as an inflaton, with mass scale $\sim 10^{13}$ GeV.  Chaotic inflation with a singlet right-handed sneutrino was proposed in \cite{Murayama:1992ua}. In \cite{Antusch:2004hd} sneutrino hybrid inflation was introduced in supergravity framework with exact R-symmetry while Heisenberg symmetry was imposed in \cite{Antusch:2008pn,Antusch:2009vg}, in order to solve the $\eta$-problem.  Non-singlet sneutrino inflation was explored in \cite{Antusch:2010va,Gonzalo:2016gey}, where the $\eta$-problem was avoided by either Heisenberg symmetry or shift symmetry.

In this work we consider a non-singlet inflaton under $U(1)_{B-L}$ gauge symmetry that is assigned to right-handed sneutrino, in no-scale supergravity. One remarkable feature of the present model is that the $U(1)_{B-L}$ gauge symmetry is broken during inflation by the inflaton hence topological defects such as cosmic strings are absent. After the inflation ends, the gauge symmetry is broken in the true minimum at GUT scale which is a derived scale from Planck scale $M_P$ and inflationary mass scale $\mu$. Moreover, the inflation fields play a crucial rule in generating tiny neutrino masses via a TeV inverse seesaw mechanism \cite{Mohapatra:1986aw,Mohapatra:1986bd,Gonzalez-Garcia:1988okv} that allows non-small values of the neutrino Yukawa couplings $Y_\nu$, without unnatural fine tuning. We introduce $U(1)_R$ symmetry (R-symmetry) on the renormalizable level while we allow non-renormaizable operators suppressed by Planck mass to break R-symmetry  \cite{Civiletti:2013cra,Khalil:2018iip}.

The paper is organized as follows. In section \ref{sec:model} we construct the model discussing the role of each supermultiplet. In section \ref{sec:inflation} we explore the inflation scenario. Also, we analyze the inflation effective potential, trajectory and observables. We discuss consequences on post inflation era in section \ref{sec:reheat-neutrino}. We point out an interesting feature of SUSY breaking in models of hybrid inflation and exist in our model. We investigate constraints that arise from reheating and then generate neutrino masses from inverse seesaw mechanism. Finally, we draw our conclusions in section \ref{sec:conclusions}.
\section{The model}
\label{sec:model}
We consider a no-scale supergravity model in which both the superpotential and the K\"ahler potential are invariant under $U(1)_{B-L}$ gauge group. Moreover, on the renormalizable level, the Lagrangian respects the $U(1)_R$ symmetry, with $R[W]=2$ while non-renormalizable operators suppressed by Planck mass are allowed to break R-symmetry \cite{Civiletti:2013cra,Khalil:2018iip}. A $Z_3$ discrete symmetry is imposed in order to guarantee a successful inflation and TeV inverse seesaw mechanism. 
The field content of the model is listed as follows.
\begin{itemize}
\item The supermultiplets $S_1=(\tilde{S}_1,S_1),\, S_2 =(\tilde{S}_2,S_2) $ are gauge non-singlets under $U(1)_{B-L}$. Their scalar components are corresponding to the inflaton while the fermionic components mix with the right-handed neutrino $N$ in the neutrino mass matrix.
\item The waterfall supermultiplets $\phi_1= (\phi_1,\tilde{\phi_1}) \,, \, \phi_2= (\phi_2,\tilde{\phi_2})$ whose scalar components are frozen at the origin during the inflation and after the inflation ends they break the $B-L$ gauge symmetry at the true minimum.

\item The singlet $S=(S,\tilde{S})$ plays a five-fold role: First, it is a driving field where its F-term generates the potential of the waterfall fields, (the scalar components of $\phi_1 ,{\phi_2}$), hence they acquire vevs to break the $B-L$ symmetry at GUT scale $M \sim \sqrt{\mu M_P}$. Second, the non-renormalizable coupling to $S_1,S_2$ will play important role in flattening the inflation potential causing wider region of the parameter space of a successful inflation as we will demonstrate in the following section. Third, its decay contributes to reheating the universe.
On the other hand SUSY breaking effects generate a tadpole term in the scalar $S$ shifting its minimum from the origin, hence the fourth role is realized in generating the mixing term $\mu_S S_1 S_2$ after the end of inflation and can contribute to the neutrino mass matrix.  Finally non-zero vev of $S$ can be employed to solve the $\mu$-problem of the TeV scale MSSM by generating the TeV scale mixing term $\mu\, H_u \,H_d$ that is needed to trigger the electroweak symmetry breaking \cite{Dvali:1997uq}. 

\item The right-handed neutrino supermultiplet $N=(\tilde{N},N)$ which contributes to the neutrino mass matrix and has no role in the inflation scenario.
  
\end{itemize}
\begin{table}[h]
 \centering
 \begin{tabular}{c | c c c c c c }
 \hline \hline
   & $\,\,\,\, S_1 \,\,\,\, $ & $ \,\,\,\,  S_2 \,\,\,\,  $ & $ \,\,\,\,  \phi_1 \,\,\,\, $ & $\,\,\,\, \phi_2 \,\,\,\,  $ & $\,\,\,\, S \,\,\,\, $ & $ \,\,\,\, N \,\,\,\, $\\
   \hline 
   $U(1)_{B-L}$ & 1 & -1 & 2 & -2  & 0 & -1\\
   $R$ & 1 & 1 & 0 & 0  & 2 & -1 \\
  $Z_3$ & $\omega^2$ & $\omega$ & $\omega$ & $\omega^2$  & 1 & $\omega$ \\
   \hline  \hline
  \end{tabular}
 \caption{{\footnotesize  $U(1)_{B-L}$ and $R$ charge assignments for different superfields $S_1,S_2,\phi_1,\phi_2,S,N$ in the inflation sector.}}
 \label{tab:B-L_R}
\end{table}
The superfields constituting the model, which are SM singlets,  as well as the respective ${B-L}$, the R-symmetry and $Z_3$ charges are given in Table ~\ref{tab:B-L_R}. In this respect the superpotential and the K\"ahler potential for the inflation sector are given by
%
%
\bea\label{eq:suppot_inf}
\!\!\!\!\!\! W_{\text inf}= \kappa_1  S \left( \phi_1  \phi_2-\mu M_P \right) + \kappa_2 \, S_1 S_2 \left( \dfrac{\phi_1  \phi_2}{M_P}-\mu \right) + \dfrac{\lambda_1}{M_P} (S_1 S_2)^2  + \dfrac{\lambda_2}{M_P} S^2 S_1 S_2,
\eea
\bea\label{eq:K1}
K=  -3 M_P^2 \log\left[\dfrac{T+\overline{T}}{M_P} - \frac{|S|^2}{3M_P^2} - \frac{|S_1|^2}{3M_P^2}- \frac{|S_2|^2}{3M_P^2}  - \frac{|\phi_1|^2}{3M_P^2} - \frac{|\phi_2|^2}{3M_P^2} \right],
\eea
The superpotential (\ref{eq:suppot_inf}) is a development of the tribrid inflation model \cite{Antusch:2004hd,Antusch:2008pn,Antusch:2009vg,Antusch:2010va,Antusch:2012jc,Antusch:2013eca}. As will be detailed later, the difference is at two main points: The first is that the mass term of right-handed neutrinos $S_1\,S_2$ vanishes at the end of inflation. The second is that the vev of $S$ is not zero during the inflation.
The parameter $\mu$ determines the scale of the inflation and $\kappa_i,\lambda_i$ are dimensionless couplings where $\lambda_i$ are responsible for the R-symmetry breaking in the superpotential and $M_P =2.4 \times 10^{18}$ GeV is the reduced Planck mass.\footnote{Worthy to mention that the symmetry allows the Planck suppressed non-renormalizable operators such as $S^4$, $ S^2 \phi_1 \, \phi_2$ and $ (\phi_1 \, \phi_2)^2$. However they are undesirable hence we can set the corresponding couplings to zero or we can argue that they are forbidden in the UV sector by $R$-symmetry. We assume that only $S,S_i,N$ as well as the MSSM superfields carry R-charges. We are not going to give more details on the UV sector as it is beyond the scope of the paper.} 
In this regard, the total scalar potential consists of the F-term and D-term parts $V=V_F+V_D$. The F-term scalar potential is given by 
\bea
V_F = e^{K/M_P^2} \left[D_I K^{I\bar{J}} D_{\bar{J}}\overline{W}- 3 \dfrac{|W|^2}{M_P^2} \right],
\eea
where $I,J$ run over $(T,S, S_1, S_2 , \phi_1,\phi_2)$, $K^{I\bar{J}}$ is the inverse of the K\"ahler metric $K_{I\bar{J}}= \dfrac{\partial K}{\partial Z^I \partial Z^{\bar{J}}}$, and $D_I$ is the K\"ahler derivative defined by $D_I=\dfrac{\partial}{\partial Z^I} + \dfrac{1}{M_P^2} \dfrac{\partial K}{\partial Z^I}$. We use the lower case letters $i,j$ to run over the inflation sector fields $S ,S_1 , S_2, \phi_1,\phi_2$.
The D-term potential is given by
\bea
V_D = \frac{g^2}{2} {\text Re}f^{-1}_{AB} D^A D^B ,
\eea
where $f_{AB}$ is the gauge Kinetic function and the indices $A,B$ are corresponding to a representation of the gauge group under which $Z^i$ are charged. The D-term $D^A$ is given by 
\bea
D^A = \frac{\partial K}{\partial Z^i} \, \left(T^A \right)^i_j Z^j,
\eea
with $T^A$ are generators of the  gauge group in the appropriate representation. We will work in the D-flat direction, hence the total potential will be given by F-term scalar potential as
\be\label{Eq:Ftermpot}
V= \dfrac{1}{\Omega^2} \, 
\mathlarger{\mathlarger{\mathlarger{‎‎\sum}}}_{i=1}^{5}\left|\frac{\partial W}{\partial Z_i}\right|^2, \hspace{1cm} \Omega =
 \frac{T+\bar{T}}{M_P}- \frac{|S|^2}{3 M_P^2} - \frac{|\tilde{S}_1|^2}{3 M_P^2} - \frac{|\tilde{S}_2|^2}{3 M_P^2} - \frac{|\phi_1|^2}{3 M_P^2} - \frac{|\phi_2|^2}{3M_P^2}.
\ee
After calculating the scalar potential, we consider the modulus $T$ to be stabilized at high scale by any of the mechanisms of \cite{Kachru:2003aw,Balasubramanian:2005zx,Ellis:2013nxa}, where $\langle {\text Re}(T)  \rangle =\tau_0$, $\langle {\text Im}(T)  \rangle =0$.
It is clear that, the total scalar potential is positive semidefinite. Therefore it has a global SUSY minimum which is Minkowskian, and is located at 
\bea
\langle S\rangle=\langle \tilde{S}_1\rangle= \langle \tilde{S}_2\rangle= 0 \,\,\& \,\,\, \langle \phi_1 \phi_2\rangle= \mu M_P
\eea 

We will work in the D-flat direction associated with the $B-L$ symmetry, 
\bea\label{eq:Dflat}
|\tilde{S}_1|^2- |\tilde{S}_2|^2 +2 |\phi_1|^2 - 2 |\phi_2|^2=0.
\eea
Without loss of generality we choose for simplicity the direction $|\tilde{S}_1|=|\tilde{S}_2|=|\Sigma|$ and $ |\phi_1|=|\phi_2|=|\phi|$, that cancels two real degrees of freedom. Therefore, the scalar potential simplifies to the following form 
\bea\label{eq:pot} 
V= \dfrac{1}{\Omega^2} \, 
\left[  \left|\kappa_1 \left(\phi^2 - \mu M_P\right) + \dfrac{2\lambda_2}{M_P} S \,\Sigma^2 \right|^2 + 2 \left| \phi \right|^2 \left| \kappa_1 S +  \dfrac{\kappa_2}{M_P} \Sigma^2  \right|^2  \nonumber \right. \\ \left.
+ 2 \left| \Sigma \right|^2  \left| \dfrac{\kappa_2}{M_P} \left(\phi^2  - \mu M_P \right) + \dfrac{\lambda_1}{M_P} \Sigma^2 
+ \dfrac{\lambda_2}{M_P} S^2 \right|^2 
\right].
\eea
The previous direction is not unique as one can choose another D-flat direction in which $|\phi_1|^2 -  |\phi_2|^2=\gamma \neq 0$, which implies that $|\tilde{S}_1|^2= { |\tilde{S}_2|^2 -2 \gamma }$, hence only one degree of freedom cancels.
However, as we will show in the next section, all the four real degrees of freedom $\alpha_{1,2},\beta_{1,2}$ are fixed to zero during inflation, where $\phi_j=\dfrac{\alpha_j +i \beta_j}{\sqrt{2}}$. Hence, during the inflation and at the global minimum $\gamma=0$, and we return to the simple case of $|\tilde{S}_1|=|\tilde{S}_2|=|\Sigma|$.
In this regard we express  the complex fields $S,\Sigma,\phi$ in terms of their real components as follows
\bea
  S= \frac{s+i\sigma}{\sqrt{2}}\,\,, \,\,\, \phi=\dfrac{\alpha +i \beta}{\sqrt{2}} \,\,, \,\,\,
  \Sigma=\dfrac{\chi +i \psi}{\sqrt{2}} .
\eea
Accordingly, the global minimum of the potential is located at
\bea
s=\sigma = \chi= \psi=\beta = 0 \,\,\& \,\,\, \alpha = \sqrt{2\mu M_P}  .
\eea
It turns out that $\chi$ and $\psi$ are massless at the true minimum, since the second term of (\ref{eq:suppot_inf}) vanishes when the higgs fields acquire non-zero vev $\sim\sqrt{\mu M_P}$. However, as we will demonstrate later, SUSY breaking effects will generate mass terms for $\tilde{S}_i$. On the other hand the squared masses of the other real components are given by 
\bea\label{eq:masses1}
m_s^2=m_\sigma^2 = \frac{3 \, \kappa_1^2 \, \mu \,  M_P^2}{3\, \tau_0 - \mu}     \,\,\,\,  \text{and} \,\,\,\, m_\alpha^2 =m_\beta^2 = \frac{ \kappa_1^2 \, \mu \,  M_P^2}{\tau_0} \,.
\eea
%
%
\section{Inflation scenario and observables}
\label{sec:inflation}
\subsection{Inflation trajectory and effective potential}
\begin{figure}[b!]
	\centering
	\includegraphics[width=0.48\textwidth]{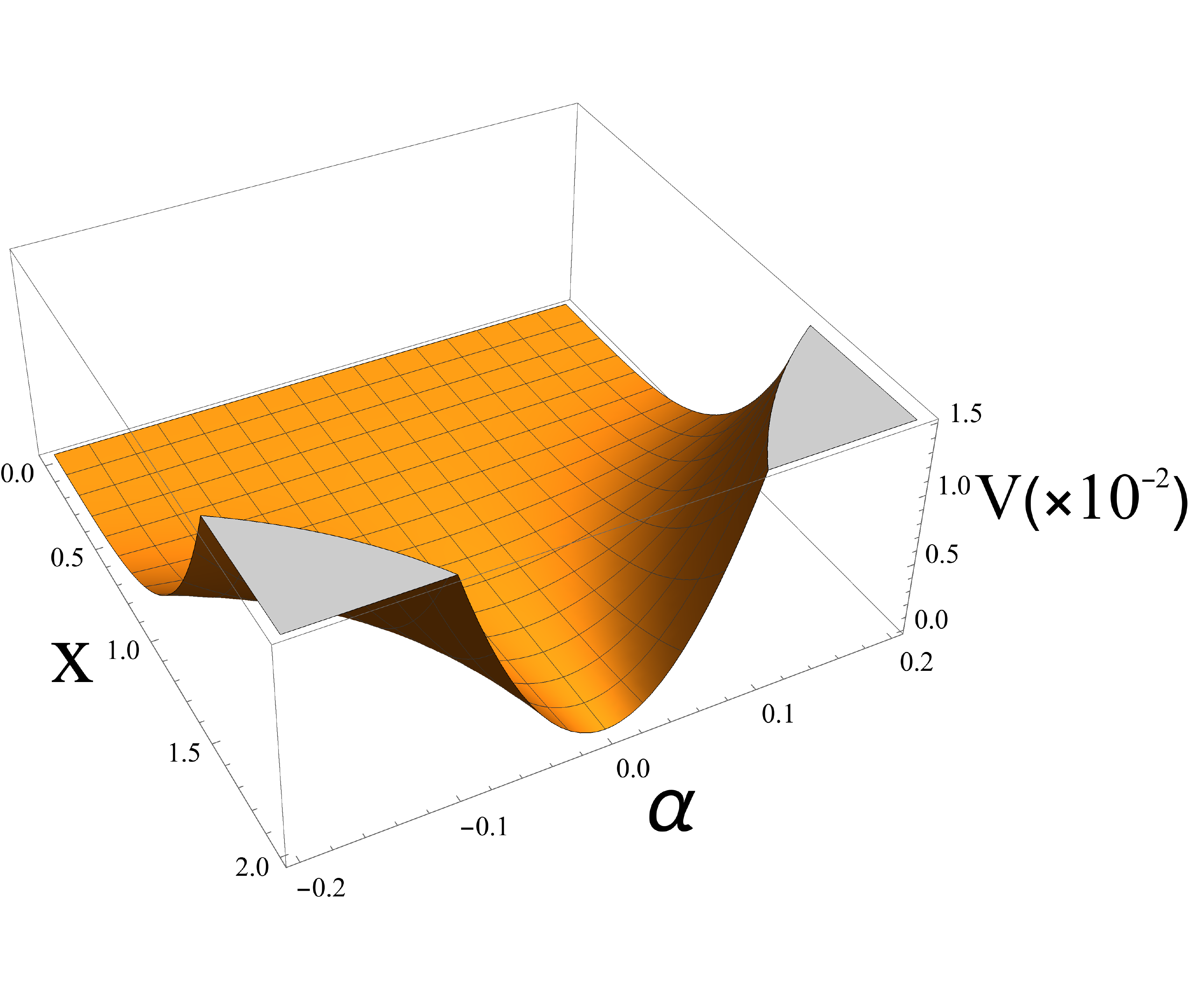}
	\hspace{0.1cm}
	\includegraphics[width=0.48\textwidth]{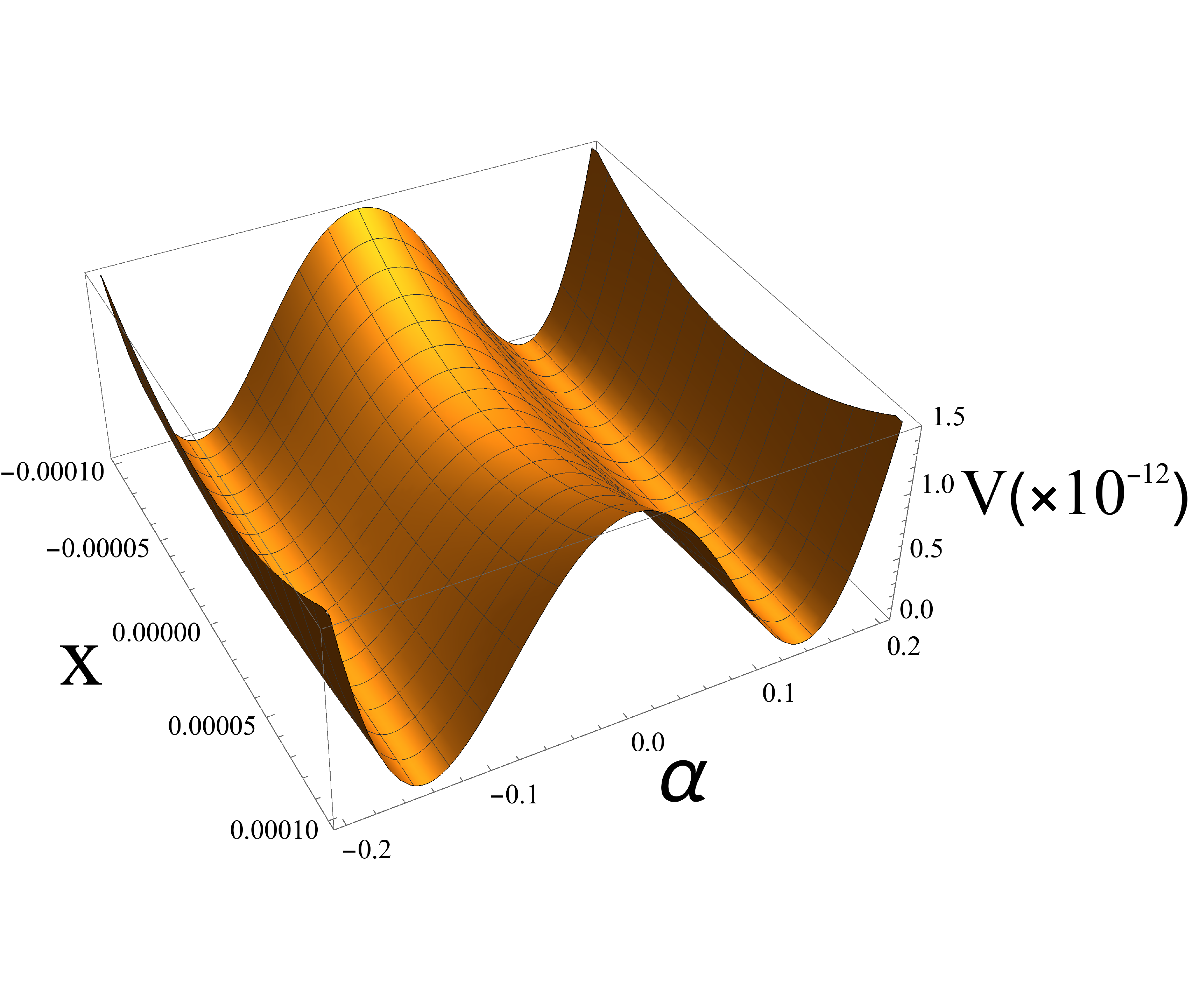}
	\caption{The scalar potential of $x,\alpha$ near the SUSY vacuum (left panel) and  for large values of $x$ (right panel), with $\beta=y=\sigma=0$ and $\lambda_1 = 0.3\times 10^{-5}, \mu= \times 10^{-5}, \kappa_1=10^{-4} , \kappa_2 \sim 1 $. All values are given in the units where $M_P=1$.
	\label{fig:3Dpot}}
\end{figure}	
Here we study the trajectory of inflation and the resulting effective inflation potential. The scalar potential is minimized at $ \phi_1 = \phi_2 = 0$ for large values of $S_1,S_2$. Along the inflation trajectory $ \phi_1 = \phi_2 = 0$ and the effective inflation potential receives a dominant contribution from the F-terms $F_{S_i}$ as well as a contribution from $F_{S}$, that will be constrained from the inflation observables, while the contribution of $F_{\phi_i}$ vanishes. On the other hand, the inflation energy shifts the vev of $s$ from zero during inflation to a small value, due to the coupling $\dfrac{\lambda_2}{M_P} S^2S_1S_2$.
We depict the shape of the scalar potential of $(x,\alpha)$ during the inflation and near the global SUSY minimum after the inflation ends in Fig. \ref{fig:3Dpot}. 

We first investigate the form of the scalar potential in the limit $\lambda_2=0$ (the potential is minimized at $S=0$), then we expand the scalar potential for small $s$ when $\lambda_2 \neq 0$. In this respect, the inflation potential, when $\lambda_2=0$, has the form
%
%
\bea
\!\!\!\!\! V(\! S \! = \! 0 \!)=\frac{9 M_P^2 \left[8 \lambda_1^2 |\Sigma|^6 + \kappa_1^2 \mu ^2 M_P^4 + 2 \kappa_2^2 \mu ^2 M_P^2 |\Sigma|^2 -4\kappa_2 \lambda_1 \mu  M_P \left(\overline{\Sigma}^4+ \Sigma ^4\right)\right]}{4  \left(|\Sigma|^2-3 M_P \tau_0\right)^2}.
\eea
In order to have canonical kinetic terms we use the following field redefinitions 
\bea\label{eq:redefin}
\Sigma=\sqrt{3M_P\tau_0}  \, \tanh\left( \dfrac{x+i y}{\sqrt{6}M_P} \right),
\eea
where $x,y$ are real scalar fields. As will be demonstrated latter, the field $y$ acquires a mass of order Hubble scale during the inflation hence it is stabilized at zero \cite{Moursy:2020sit}. Therefore, the scalar potential takes the form
\bea\label{eq:pots0}
\!\!\!\!\!\! V=  A \cosh ^4\left(\frac{x}{\sqrt{6}}\right) \left[B^2 \tanh ^6\left(\frac{x}{\sqrt{6}}\right)-2 B \tanh ^4\left(\frac{x}{\sqrt{6}}\right)+\tanh ^2\left(\frac{x}{\sqrt{6}}\right)  +f \right] \! ,
\eea
where, 
\bea
A= \dfrac{3\mu^2 \kappa_2^2}{2\tau_0 M_P}\, , \hspace{0.5cm} B=\dfrac{6\lambda_1\tau_0}{ \kappa_2 \,\mu}\, , \hspace{0.5cm} f=\dfrac{\kappa_1^2 M_P}{6\kappa_2^2\tau_0} ,
\eea
 are dimensionless parameters that will be determined by slow-roll conditions and inflation observables. We will use the units in which $M_P=1$ throughout this section. When $f=0$ (or equivalently $\kappa_1=0$), we obtain the potential studied in \cite{Khalil:2018iip}. Therefore, in the limit where $B=1$ and $f=0$, the potential (\ref{eq:pots0}) is asymptotically flat 
\bea\label{eq:pot0}
V_{0}= A \tanh^2\left(\dfrac{x}{\sqrt{6}}\right).
\eea

Now we return to consider the effect of the coupling $\dfrac{\lambda_2}{M_P} S^2S_1S_2$. It gives rise to a linear term in $s$ in the $F_S$ contribution to the potential (\ref{eq:pot}), hence $s$ acquires small vev during the inflation and it is given by 
\bea
s=\frac{\sqrt{2 A f}}{3 \lambda _2},
\eea
where we have expanded for large $x$ during the inflation and taken the leading order.\footnote{This will not affect the canonical kinetic term of the inflaton as $s$ is much small during inflation.} 
\begin{figure}[t!]
	\centering
	\includegraphics[width=0.6\textwidth]{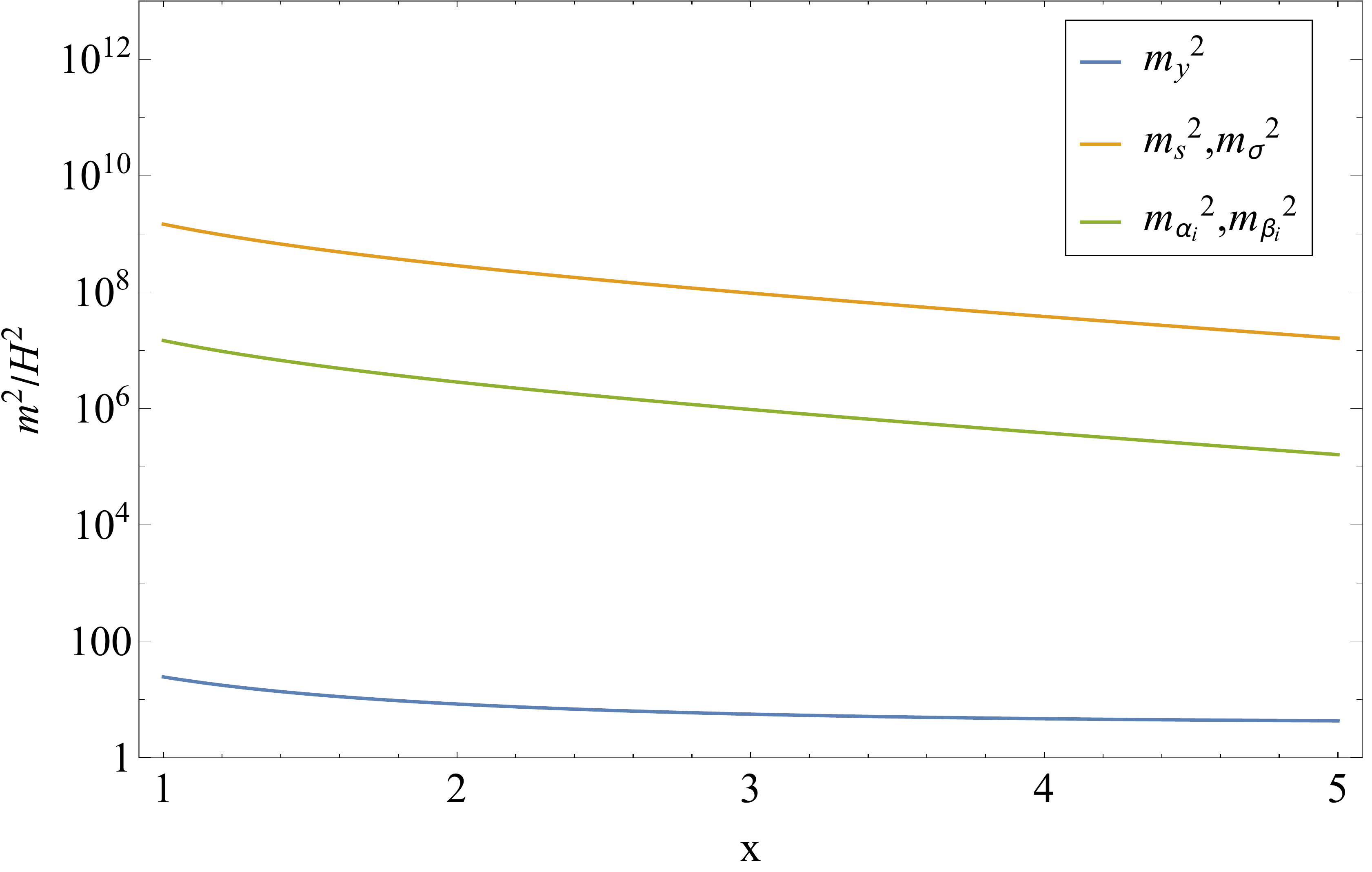}
	\caption{The ratio of the squared mass of the real scalar fields $\alpha_i,\beta_i, y$ and $\sigma$, to the squared Hubble scale, with $A=3\times 10^{-10}, f=3.33\times 10^{-11} \,,\, \lambda _2 = 0.005,\, B=1$. All values are given in the units where $M_P=1$.	 
	\label{fig:mass2}}
\end{figure}	

Now, we consider for completeness, that all real components of $\phi_{1,2}$ that can account for possible D-flat directions, $\alpha_i,\beta_i$. We show that they are fixed at zero during inflation as well as the real degrees of freedom $y,\sigma$. 
 As a matter of fact, all real degrees of freedom, apart from the inflaton $x$, acquire masses larger than the Hubble scale $H$ during inflation \cite{Moursy:2020sit}. We present the ratio of field dependent squared masses to the squared Hubble scale during inflation in Fig. \ref{fig:mass2}. 
 
 Returning to the scalar $\alpha$ that accounts for the waterfall phase,  the scalar potential (\ref{eq:pot}) is minimized at $\alpha=0$  during the inflation, for $x$ is rolling at large values, since the field dependent squared mass  $m^2_\alpha$ is positive and is of order squared Hubble scale during inflation. When $x$ reaches a critical value 
$x_c$ after the inflation ends, $m^2_\alpha$ flips it sign to negative values which triggers the waterfall phase \cite{Moursy:2020sit}.
Expanding for small values of the inflaton $x$, we have 
\bea
m_{\alpha }^2\simeq \frac{\kappa _1^2 \,\mu  \left(\mu -3 \tau _0\right)}{6 \tau _0^2},
\eea 
which is negative as long as $\mu < 3 \tau _0$. The critical value of the inflaton $x_c$ will be given as
\bea
x_c=\frac{2 \kappa _2 \mu }{\tau_0\, \left(\kappa _2+4 \lambda _1\right)}.
\eea
\begin{figure}[t!]
	\centering
	\includegraphics[width=0.6\textwidth]{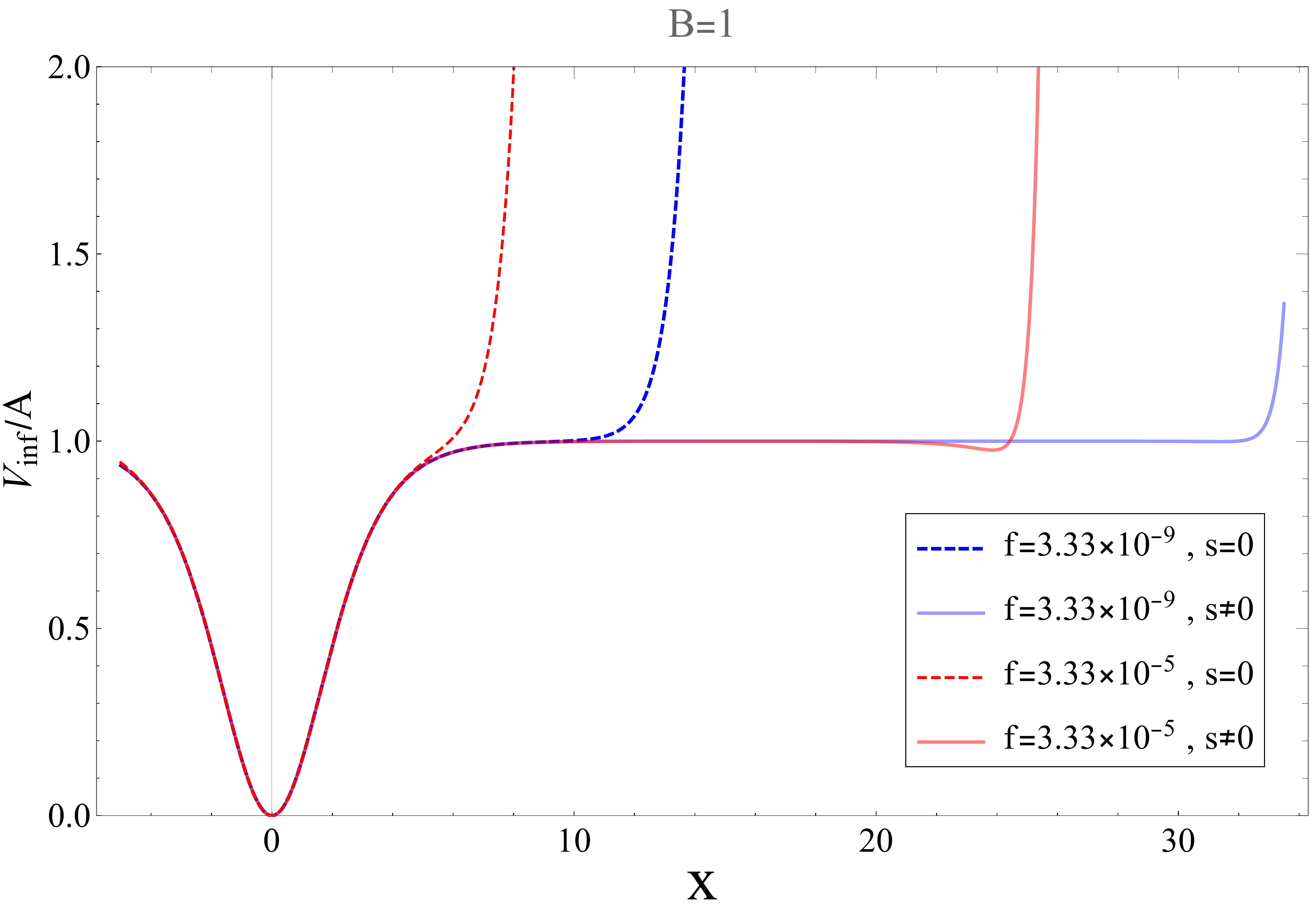}
	\caption{The scalar potential $\dfrac{V_{\text{inf}}}{A}$ of the inflaton $x$ in the two cases when $s$ is zero (solid curves) and non-zero (dashed curves), with $A=3\times 10^{-10} \,,\, \lambda _2 = 0.5,\, B=1$ and we took two values for $f=3.33\times 10^{-9}$ and $3.33\times 10^{-5}$. All values are given in the units where $M_P=1$.
	\label{fig:pot1}}
\end{figure}	

We expand the scalar potential (\ref{eq:pot}) to second order in $s$ with the other fields being fixed at the origin, and consider the limit where $B=1$, then we obtain the inflation effective potential as follows
\bea\label{eq:potexp1}
\!\!\!\!\!\!\!\!\! V_{\text{inf}} &=&  V_0 + A \,f \cosh ^4\left(\frac{x}{\sqrt{6}}\right) - 6 \, \lambda_2 \, \sqrt{ \dfrac{f}{2\,A}}   \, s \, V_0 \, \cosh ^4\left(\frac{x}{\sqrt{6}}\right)  + \frac{s^2 \cosh ^2\left(\frac{x}{\sqrt{6}}\right)}{6 A \tau _0} \times \nonumber\\
&&
\times \left[A \, f \cosh ^4\left(\frac{x}{\sqrt{6}}\right)+V_0 \left(-3  A\, \lambda _2 \sqrt{\frac{6\tau _0}{A}}+A+27 \lambda _2^2 \,\tau _0\, \sinh ^2\left(\frac{x}{\sqrt{6}}\right)\right)\right]\!\! .
\eea
\begin{figure}[b!]
	\centering
	\includegraphics[width=0.48\textwidth]{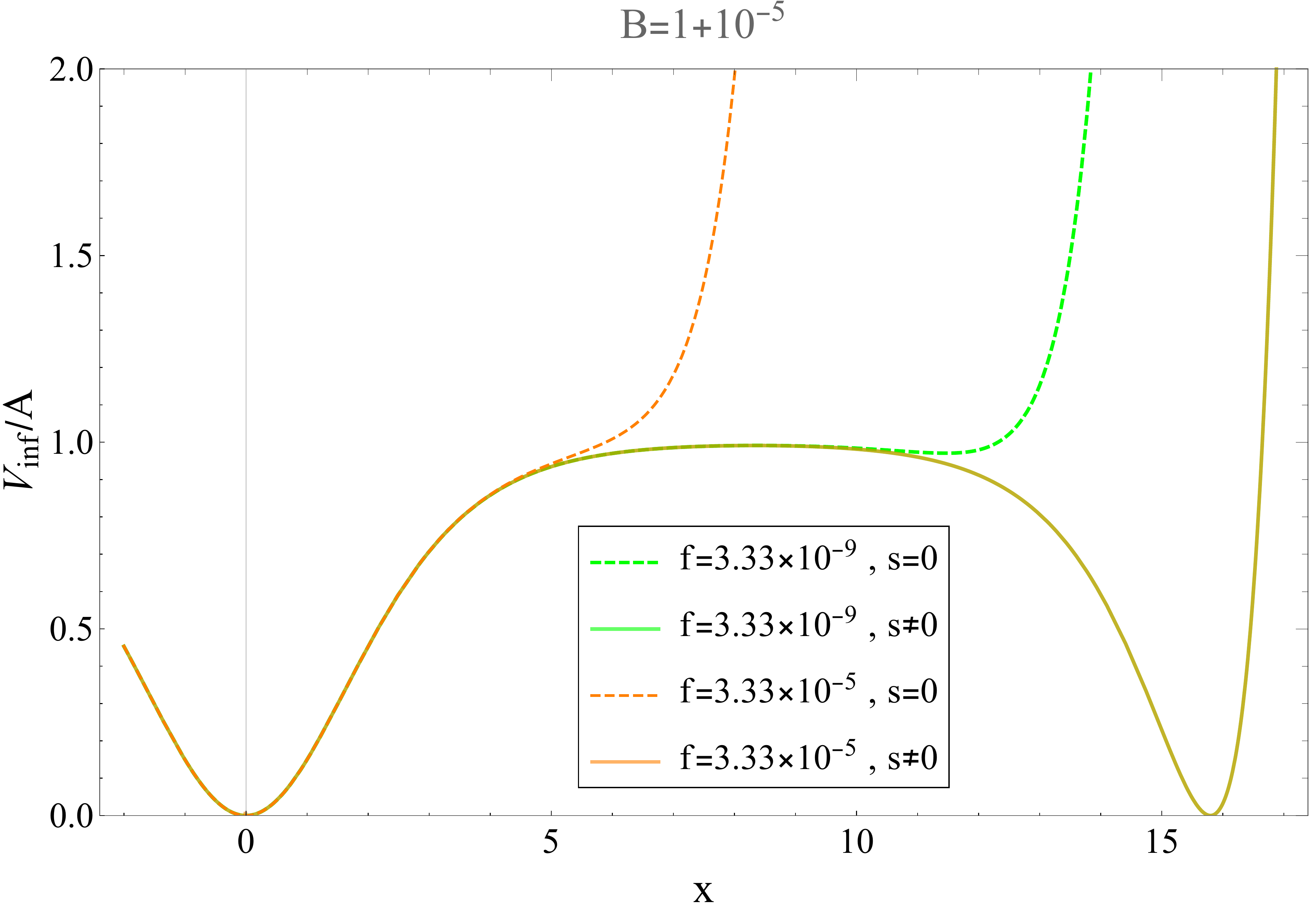}
	\hspace{0.2cm}
	\includegraphics[width=0.48\textwidth]{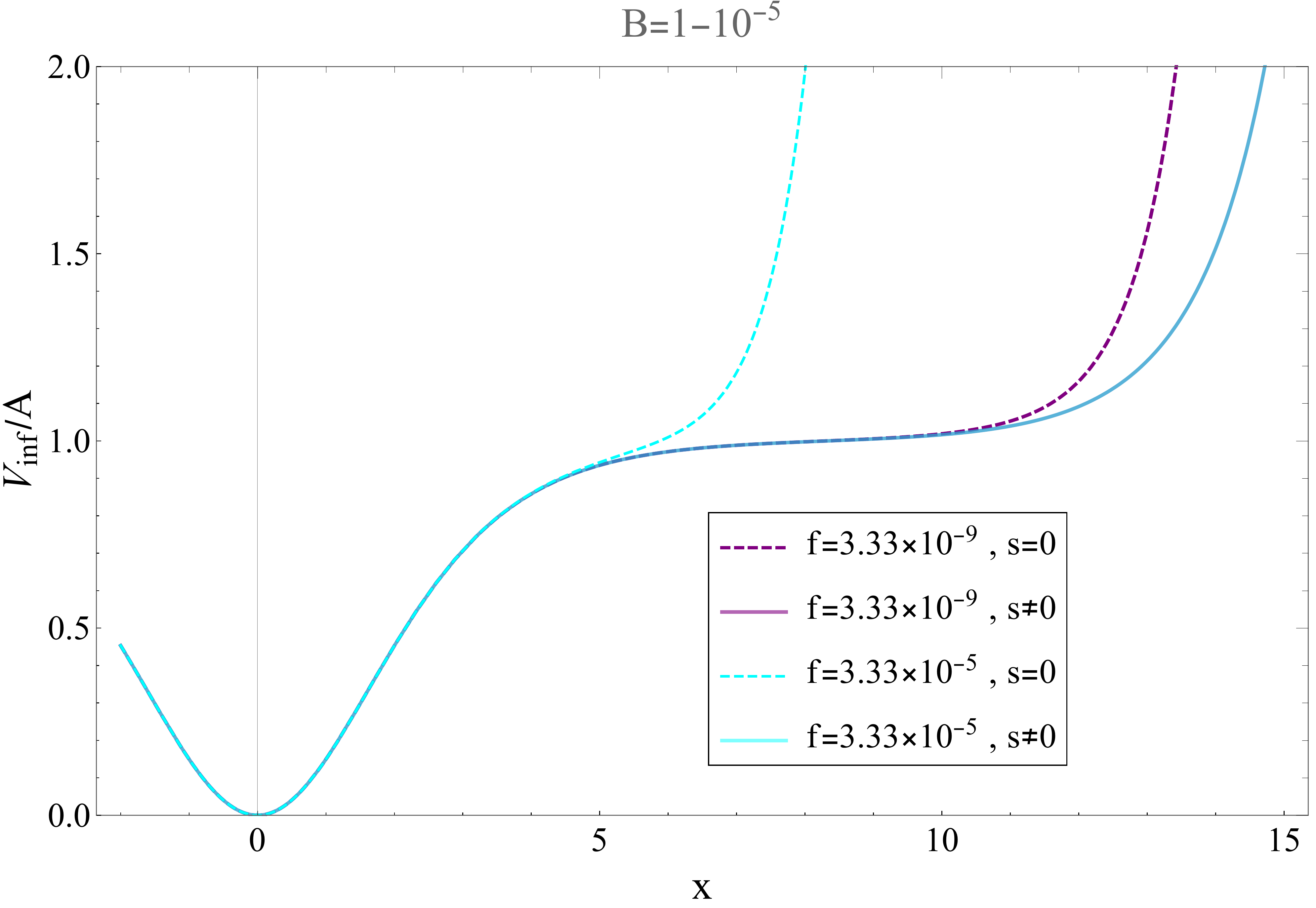}
	\caption{The inflation potential $\dfrac{V_{\text{inf}}}{A}$ with $A=3\times 10^{-10},\lambda _2=0.5$, $s\neq 0$ and $s=0$ for the solid and dashed curves respectively, and we took two values for $f=3.33\times 10^{-9}$ and $3.33\times 10^{-5}$.
The left panel corresponds to $B=1+10^{-5}$ and the right panel corresponds to $B=1-10^{-5}$. All values are given in the units where $M_P=1$.
	\label{fig:pot2-3}}
\end{figure}	
If $\lambda_2=0$, then $s$ will be stabilized at zero during the inflation, hence the third and fourth terms will be absent in the inflation potential (\ref{eq:potexp1}), and we return to the potential (\ref{eq:pots0}). It is clear that the second term proportional to $f$ spoils the asymptotic flatness of the potential at large values of the inflaton $x$, hence $f$ should be small enough in order to guarantee the flatness of the potential during the inflation. For small values of $s$, the third term dominates over the fourth term during the inflation. As a consequence, the third term in (\ref{eq:potexp1}) has a desirable effect in cancelling the dangerous effect of the second term hence flattening the potential during the inflation. In Fig. \ref{fig:pot1} we show the inflation potential for $B=1$ when $s$ is stabilized at zero and non-zero values during inflation where it was indicated that the potential is flattened for $s\neq 0$ during the inflation as depicted in the solid curves. When $B\neq 1$, we define $B(\xi\neq 0)=1+\xi $, hence extra terms proportional to $\xi$ are generated and this may spoil the flatness during the inflation. Accordingly the constraint  $|\xi| \lesssim {\cal O}(10^{-4})$ should be imposed \cite{Khalil:2018iip}. Fig. \ref{fig:pot2-3} illustrates the inflation potential for $B\neq 1$ ($\xi\neq 0$) for different values of $f$. The solid curves corresponding to non-zero values of $s$, which coincide for different values of $f$. 
%
\subsection{Inflation observables}
Now we move to discuss the model predictions of inflation observables and explore the possible constraints on the different parameters.
The inflation observables, tensor-to-scalar
ratio $r$, spectral tilt $n_s$ and the scalar amplitude $A_s$ will be investigated. In terms of the slow-roll parameters $\epsilon$ and $\eta$, the observables are defined as follows
\bea
r &=& 16 \epsilon , \nonumber \\
n_s&=& 1-6 \epsilon + 2 \eta , \nonumber \\
A_s &=& \frac{V}{24 \pi^2 \epsilon},  \nonumber 
\eea
where the above observables are computed at the crossing horizon value of the inflaton field $x_*$. The number of e-folds is given by
\bea
N= \int_{x_e}^{x_*} \frac{1}{\sqrt{2 \epsilon}} \, dx,
\eea
where $x_e$ is the value of the inflaton at the end of inflation.
The observed value of the scalar amplitude $A_s\simeq 1.95896 \pm 0.10576 \times 10^{-9}$ at 68\% CL  \cite{Akrami:2018odb}, which fixes  $A \sim 10^{-10}$.

\begin{figure}[h!]
	\centering
	\includegraphics[width=0.65\textwidth]{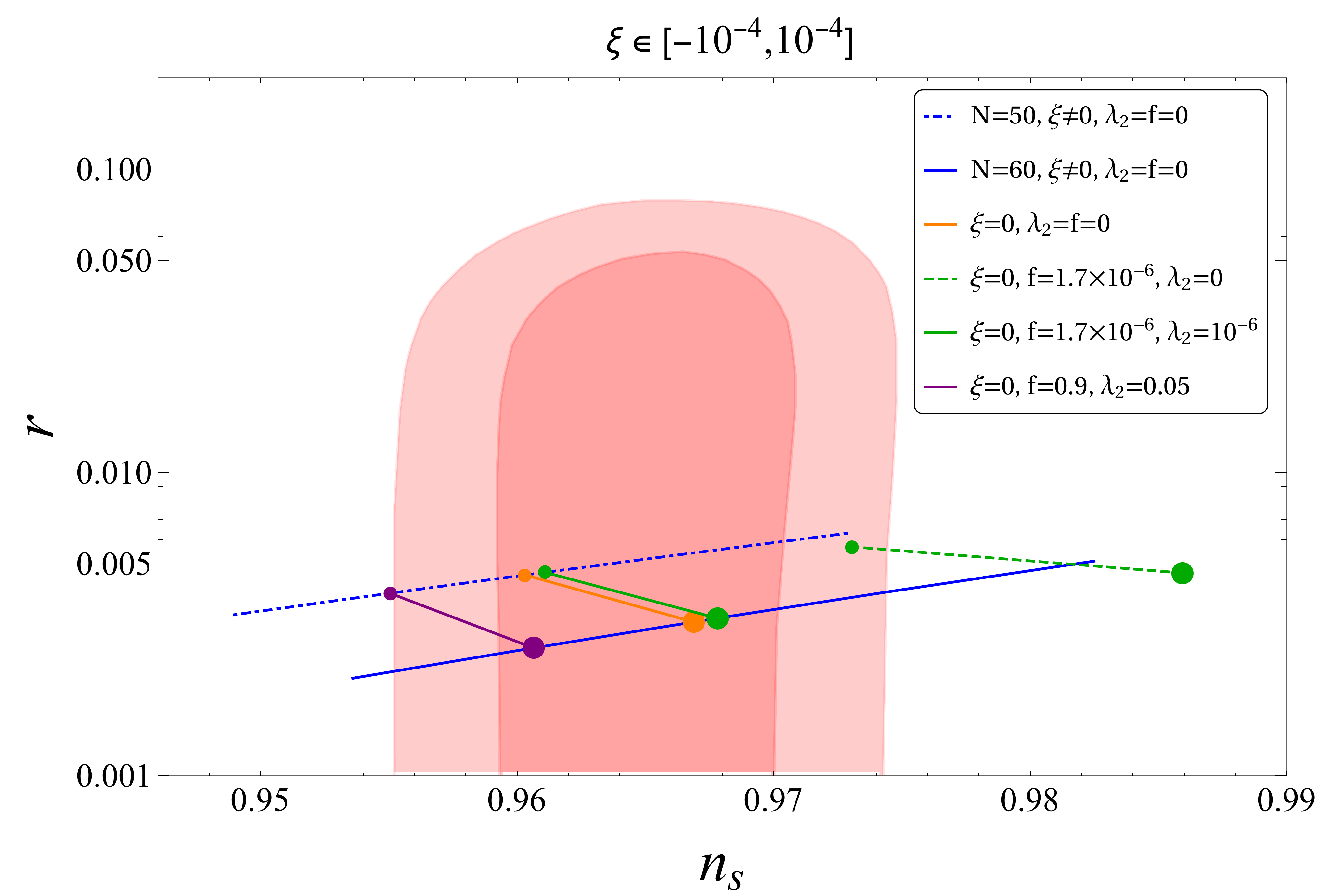}
	\caption{A logarithmic plot for the model predictions  in the $n_s-r$ plane. We have fixed $A\sim 10^{-10}$. The solid (dash-dotted) blue curves correspond to $N =60 $ ($N = 50$) respectively where we scanned over $B=1+\xi$ with $\xi \in [-10^{-4} , 10^{-4}]$, with $f=\lambda_2=0$. The other segments correspond to fixed $B=1 \,(\xi=0)$ with different values of $f,\,\lambda_2$ and we have scanned over $N=50-60$ (small-big) dots, for each segment. The dark and light red regions correspond to the 1 and 2 sigma exclusion limits released by the Planck collaboration (2018)
 (TT+ TE+ EE + LowE + Lensing + BK14)\cite{Akrami:2018odb}.
	\label{fig:planck1}}
\end{figure}	

In Fig. \ref{fig:planck1}, we displayed the predictions of our model for $n_s$, $r$ in a logarithmic plot versus the Planck limits.
First, we fixed $f=\lambda_2=0$  and scanned over $B=1+\xi$ with $-10^{-4} \, \leq \, \xi \, \leq \, 10^{-4}$ on the solid (dash-dotted) blue curves for $N =60 $ ($N = 50$), respectively. 
On the other segments, we scanned over $N=50-60$, and fix $B=1 \, (\xi=0)$. The orange segment illustrates the situation when the  potential is asymptotically flat (\ref{eq:pot0}) where $f=\lambda_2=0$. However, $f$ (or $\kappa_1$) should not be zero in order to keep the potential terms responsible for stabilizing the waterfall fields at the true minimum. Therefore fixing $\lambda_2=0$ (which means that $s=0$ on the inflation trajectory), the inflation observables constrain $f\lesssim 2\times 10^{-6}$ as indicated in the green dashed segment. 
Turning on $\lambda_2$ to non-zero value will flatten the potential causing successful inflation. The latter case is illustrated in the solid green segment where we took the same value of $f$ on the dashed one and let $\lambda_2 \sim 10^{-6} $. Any greater value of $\lambda_2$ makes the potential more flat at larger inflaton value. Therefore we investigate greater values of $f\lesssim {\cal O}(1)$ that spoil the flatness of the potential against greater values of $\lambda_2$. The Purple segment corresponds to $\lambda_2 \sim 0.05$ and $f\sim 1$. For larger value of $\lambda_2$ purple segment moves towards the orange one. For $\lambda_2 \gtrsim 0.5$, the purple and orange segments coincide. For $B\neq 1$ and  $\lambda_2 \gtrsim 0.5$, different potentials with different values of $f$ coincide as depicted in Fig. (\ref{fig:pot2-3}). In order to have successful inflation, the constraint $|\xi| \lesssim {\cal O}(10^{-4})$ should be satisfied.
%
\begin{table}[h]
 \centering
 \begin{tabular}{c c | c c c c c c }
 \hline \hline
  $\,\,\,\, \lambda_2 \,\,\,\, $ & $ \,\,\,\,  f \,\,\,\,  $ & $ \,\,\,\, \mu \,\,\,\, $ & $ \,\,\,\,  \kappa_1 \,\,\,\, $ & $\,\,\,\, \kappa_2 \,\,\,\,  $ & $\,\,\,\, \lambda_1 \,\,\,\, $ & $\,\,\,\,\tau_0 \,\,\,\,$  & $\,\,\,\, M \,\,\,\,$ \\
   \hline 
    \hline 
   $0-10^{-6}$ & $1.7 \times 10^{-6}$  &   \begin{tabular}{@{}l@{}}
                   $2.8 \times 10^{-6}$\\
                   $2.8\times 10^{-5}$\\
                 \end{tabular} &   \begin{tabular}{@{}l@{}}
                   $10^{-3}$\\
                   $10^{-2}$\\
                 \end{tabular} &  $0.9$  &  \begin{tabular}{@{}l@{}}
                   $4.3 \times 10^{-6}$\\
                   $4.3 \times 10^{-7}$\\
                 \end{tabular} &  \begin{tabular}{@{}l@{}}
                   $0.1$\\
                   $10$\\
                 \end{tabular} & \begin{tabular}{@{}l@{}}
                   $1.6 \times 10^{-3}$\\
                   $5.3 \times 10^{-3}$\\
                 \end{tabular}\\
   \hline 
    $0.05-0.9$ & $0.9$  &   \begin{tabular}{@{}l@{}}
                   $4 \times 10^{-6}$\\
                   $3.6 \times 10^{-3}$\\
                 \end{tabular} &   \begin{tabular}{@{}l@{}}
                   $0.95$\\
                   $0.1$\\
                 \end{tabular} &  \begin{tabular}{@{}l@{}}
                   $0.9$\\
                   $0.01$\\
                 \end{tabular} &  \begin{tabular}{@{}l@{}}
                   $3 \times 10^{-6}$\\
                   $3 \times 10^{-7}$\\
                 \end{tabular} &  \begin{tabular}{@{}l@{}}
                   $0.2$\\
                   $20$\\
                 \end{tabular} & \begin{tabular}{@{}l@{}}
                   $2 \times 10^{-3}$\\
                   $6 \times 10^{-2}$\\
                 \end{tabular}\\   
   \hline  \hline
  \end{tabular}
 \caption{{\footnotesize The numerical values of the parameters of the superpotential and the modulus vev $\tau_0$ in the units $M_P=1$. We have considered $B(\xi=0)=1$ and $A\sim 10^{-10}$.}}
 \label{tab:numparam}
\end{table}
In Table \ref{tab:numparam}, we have provided some specific values of the superpotential parameters as well as the modulus vev $\tau_0$ corresponding to the values of $A,f,\lambda_2$ and $B=1$, that were discussed above in the inflation observables.
 We have focused on showing different regimes of small and large values of $\kappa_1$ and $\kappa_2$, since they play important roles in the inverse seesaw mechanism that we will demonstrate in the next section. We emphasis that $f$ can't be zero in order to account for the correct dynamics of the waterfall fields.
\subsection{Loop corrections}
Here we discuss the effects from one-loop correction to the inflation potential, and two loop correction contributing to the gauge $\eta$-problem (or Dvali problem) mentioned in \cite{Dvali:1995fb}. We will show that they don't have significant effect on the inflation potential.
\begin{itemize}
\item  The Coleman-Weinberg one-loop corrections \cite{Coleman:1973jx} are given by
\begin{equation}
 V_{\text{1-loop}}(\Sigma) = \frac{1}{64\pi^2} \text{Str} \left[m^4 \left(\log\left(\frac{m^2}{Q^2}\right) - \frac{3}{2}\right) \right],
\end{equation}
where $\text{Str} \, m(\Sigma)^4 = \, \sum_i (-1)^{2J_i} (2J_i +1 ) m_i^4 $ is the supertrace taken over all inflaton dependent mass eigenvalues of states with spin $J_i$. The stabilized fields during the inflation have $m_i \sim H$. But $H^2 M_p^2\sim V_{\text{inf}}\sim A \,M_p^4$, therefore $H^2 \sim 10^{-10} \,M_P^2$ during the inflation. Accordingly, the 1-loop correction $V_{\text{1-loop}} \sim \frac{H^4}{64 \pi^2} \sim {\cal O}(10^{-22}) \, M_P^4$, which is negligible compared to the tree level potential for any relevant renormalization scale.

\item For the two loop Dvali problem, it may arises in the context of hybrid inflation if there is a gauge non-singlet flat directions ($\Sigma$ in our case) as well as a singlet $S$ coupled to the non-singlet waterfall fields. This results in large vacuum energy by the F-term $F_S$ that breaks SUSY during the inflation. Therefore a significant two loop contribution to $\Sigma$ effective mass is estimated as \cite{Antusch:2010va}
\bea
\delta m^2 \sim \dfrac{g_{B-L}^4}{(4\pi)^4} \dfrac{|W_S|^2}{m_\phi^2},
\eea
where $m_\phi$ is the SUSY preserving mass of the waterfall fields and $g_{B-L}$ is the ${B-L}$ gauge coupling. If $\delta m \sim H$, this implies $|\eta|\sim 1$ and  the slow-roll conditions are violated causing an $\eta$-problem called the "gauge $\eta$-problem" or "Dvali Problem". However we will show that the two loop contributions are suppressed compared to the tree level inflaton mass during inflation. Similar to the situation in \cite{Gonzalo:2016gey}, SUSY breaking scale $M_{SB}$ during inflation gets contributions from $F_S= \kappa_1\mu M_P$ and $F_{S_i}$. However the dominant contribution comes from $F_{S_i}$.
According to the non-renormalization theorem, corrections should be proportional to powers of $M_{SB} \sim \langle F_{S_i} \rangle$. Therefore the corrections arise from the following contributions \cite{Gonzalo:2016gey,Antusch:2010va},
\bea
\delta m_{\Sigma}^2 \sim \,\,\, \dfrac{g_{B-L}^4}{(4\pi)^4} \dfrac{M_{SB}^4}{M_{Z'}^2}\,, \hspace{0.5cm}
\dfrac{g_{B-L}^4}{(4\pi)^4} \dfrac{M_{SB}^4 H}{M_{Z'}^2}\,, \hspace{0.5cm}
\dfrac{g_{B-L}^4}{(4\pi)^4} \dfrac{M_{SB}^4 H^2}{M_{Z'}^2}\,, 
\eea
where $M_{Z'}\sim g_{B-L} \langle \Sigma \rangle $ is the $U(1)_{B-L}$ gauge boson. Assuming that during inflation, $\Sigma \sim 1 M_P$, 
$g_{B-L} \sim 0.1$, $\mu\sim 10^{-5} M_P$, $\kappa_1\leq 0.1$, $\kappa_2\sim {\cal O}(1)$, $\lambda_1\sim 10^{-5} $, $\tau_0\sim 1 M_P $, we find 
\bea
\delta m_{\Sigma}^2 \sim \,\,\,  10^{-14} \, M_P^2 \,, \hspace{0.5cm}
  10^{-19} \, M_P^2  \,, \hspace{0.5cm}
  10^{-24} \, M_P^2    \,, 
\eea
which are negligible compared to the tree level mass squared of the inflaton during the inflation $m_{\Sigma}^2 \sim \mu^2 \sim 10^{-10} M_P^2$. Hence Dvali problem is absent in our scenario.
\end{itemize}
%
\section{SUSY breaking, reheating and neutrino masses}
\label{sec:reheat-neutrino}

In this section we investigate the link between the inflation sector and low energy physics sectors. We study the stringent constraints arising from reheating scenarios on the parameter space that is allowed by inflation observables as demonstrated in the previous section. Also, we discuss the neutrino masses which are one of the important motivations of our model. They are connected to the inflation sector via $S,\,S_1,\, S_2$. As we will see, SUSY breaking and $R$-symmetry will play essential rule in providing a tiny mass term for $S_1\, S_2$, hence TeV inverse seesaw mechanism can be realized, with $Y_\nu$ not being small. 

The assigned $B-L$ and $R$ charges in Table \ref{tab:B-L_R}, forbid the terms $S N N $, $\phi_1 N N $ as well as higher order operators containing $NN$.\footnote{Here operators such as $ (\phi_1\phi_2)^n \phi_1 N N $ and $(\phi_1\phi_2)^n \,N\, S_1$ are not allowed by R-symmetry in the UV sector as we indicated in Section \ref{sec:model}. On the other hand the operator $ S^2 \phi_1 N N $ allowed by the symmetry will be suppressed by $M_P^2$ and has no effect in the neutrino mass matrix.} 
Therefore the Majorana mass term ${M_N NN}$ is suppressed. 
 On the other hand $Z_3$ symmetry with the assigned charges in Table ~\ref{tab:B-L_R}, prevents terms such as $\phi_1 S_2^2$ and $\phi_2 S_1^2$ hence any operator containing $ S_1^2$ and $ S_2^2$ is forbidden. The relevant terms in the superpotential to the neutrino masses which are allowed by symmetry, are given by 
\bea\label{eq:suppot_nu}
\!\!\!\!\!\!\!\!  W_{\nu}= Y_\nu L H_u N+ Y_S \, S\,N\, S_1 + \dfrac{\lambda_3}{M_P} \,S \, \phi_1 \, N \, S_2 + \left[ \dfrac{\lambda_2}{M_P}S^2+ \kappa_2  \left( \dfrac{\phi_1  \phi_2}{M_P}-\mu \right)\right]S_1 S_2    ,
\eea
where the last term is common between the inflation and neutrino sector, hence a clear connection between the inflation and low energy physics of neutrino arises. SUSY breaking effects generate a tadpole term in the scalar potential $-2 \kappa_1 M^2 \, m_{3/2} \, S + h.c.$, causing  a shift in the vev of $S,\phi_1,\phi_2$ in terms of the gravitino mass $m_{3/2}$, as indicated in \cite{Dvali:1997uq,Barbieri:1982eh,Buchmuller:2000zm}
\bea\label{eq:vev-shift}
\langle S \rangle \simeq \dfrac{m_{3/2}}{\kappa_1} \,\,, \hspace{1cm}
\langle |\phi_1| \rangle = \langle |\phi_2| \rangle  \simeq M \left(1-  \dfrac{m_{3/2}^2}{\kappa_1^2 M^2} \right).
\eea
%
As advocated in the previous section, the term proportional to $S^2$ in the superpotential (\ref{eq:suppot_nu}) had an important role in flattening the inflation potential, and here it appears the third and fourth roles of $S$ in reheating and  generating the tiny neutrino masses. Indeed a mass term $\mu_S \,S_1 \,S_2$ will be induced due to the shifts in the vevs of $S$ and $\phi_i$ in (\ref{eq:vev-shift}). 
\subsection{Constraints from reheating scenarios}
\label{subsec:reheat}
At early time, after the inflation ended, the inflaton fields $\tilde{S}_1 , \tilde{S}_2$ are massless. However, they acquire masses $m_{S_i}\ll M$, due to SUSY breaking. Therefore reheating the universe occurs via the decay of the heavy inflation sector fields $S$ and $\phi_{i}$ to $N(\tilde{N})$ and $ S_i(\tilde{S}_i)$, according to the interactions in the superpotential (\ref{eq:suppot_nu}). Moreover $\tilde{S}_1 , \tilde{S}_2$ can decay to the SM particles and contribute to the reheating according to their masses. In this respect the reheating temperature is given as \cite{Giudice:2003jh,Pradler:2006hh,Ellis:2015jpg,Ellis:2016spb,Dudas:2017rpa,Ellis:2020lnc}
\bea\label{eq:TR}
T_R = \left(\frac{10}{g_s}\right)^{1/4}\left( \frac{2 \, \Gamma\,  M_P}{\pi \, c} \right)^{1/2},
\eea
where $\Gamma$ is the total decay width of the inflation sector fields, $c$ is a constant of $\cal O$(1), and $g_s$ is the effective number of light degrees of freedom in the thermal bath  at $T_R$. 

In models of supergravity inflation, an upper bound on the reheating temperature arises due to the gravitino overproduction problem \cite{Ellis:1982yb,Khlopov:1984pf,Moroi:1995fs,Ellis:1984eq,Ellis:1984er,Moroi:1993mb,Kawasaki:2004yh, Antusch:2010mv, Kawasaki:2008qe,Ferrantelli:2010as}. Gravitinos can be produced in the early universe thermally or non-thermally. If gravitinos are the lightest supersymmetric particle (LSP), then they are stable and can be a dark matter candidate. However, for high $T_R$, the  energy density of gravitinos can be larger than the present total energy density of the universe hence overcloses the universe. Since the number density of gravitinos is proportional to $T_R$, this leads to a bound on the reheating temperature in terms of $m_{3/2}$. On the other hand if gravitinos are not stable, they may decay into the LSP (which can be neutralinos) before the Big
Bang Nucleosynthesis (BBN), and hence the dark matter abundance imposes a constraint on gravitino production. Moreover, they can decay during and after the BBN. Constraints from BBN on both stable and unstable gravitinos were discussed in \cite{Kawasaki:2008qe}. If $m_{3/2} \lesssim 10$ TeV, most of gravitinos decay after the start of BBN. Therefore, preserving the BBN predictions, which is compatible with the observed abundances of the light elements $^3$He, $^4$He, $^6$Li and D, requires small abundance of gravitinos. This  imposes an upper bound on $T_R$ depending on the MSSM mass spectrum and $m_{3/2}$ as follows
\bea
 10^6 \, \text{GeV} \lesssim T_R \lesssim 2\times 10^{10}\, \text{GeV} \, , \hspace{0.5cm} {\text{for}} \hspace{0.5cm}  300 \, \text{GeV}\, \lesssim m_{3/2} \lesssim 3\times 10^{3} \, \text{GeV} \,.
\eea
 For stable gravitinos, the MSSM-LSP becomes the next to LSP (NLSP) that decays to gravitinos. If their life time is long enough, this may affect the abundances of the light elements. Therefore BBN constrains $m_{3/2}$.
 
Nonetheless, the above derived upper bounds on reheating temperature can be relaxed to $\sim {\cal O}(10^{14})$ GeV as indicated in \cite{Ferrara:2010in,Ellis:2016spb}, where  models of higgs inflation usually predict such a high $T_R$. It was pointed out that the gravitino production problem can be addressed if gravitinos are  heavy enough  $ m_{3/2} > {\cal O} $(30 TeV) such that they decay before the BBN,  or very light as in gauge mediation SUSY breaking models, $m_{3/2} \ll 1$ GeV where an upper bound on $T_R \lesssim {\cal O} (10^{14} \, \text{GeV})$ is not problematic with the BBN observations.

Now we discuss the constraints from the reheating scenario imposed on the model parameters. 
We extract the interactions of $S_i$, $S$ and $\phi_{i}$ leading to decay into fermions and scalars, that are not suppressed and contribute to $T_R$, as follows 
\bea\label{Eq:reheat_S1,2}
{\cal L}_{S_i} = \left( Y_S Y_\nu^*\frac{m_{3/2}}{\kappa_1} \, S_1 + \lambda_3 Y_\nu^* \, \frac{M \,m_{3/2} }{\kappa_1 M_P} \, S_2\right)\, \tilde{L}^* H_u^*+ \, h.c. 
\eea 
\bea\label{Eq:reheat_phi}
{\cal L}_{\phi_i} &=&  \kappa_2^2 \,\mu \, \sqrt{\dfrac{\mu}{M_P}} \, (\phi_1+ \phi_2 )(\tilde{S_1} \tilde{S}_1^* + \tilde{S_2} \tilde{S}_2^*)
 + \kappa_2\sqrt{\dfrac{\mu}{M_P}} \left( \phi_1 + \phi_2 \right) \, \bar{S_1}^c \, S_2^c \, + \, h.c. 
\eea 
\bea\label{Eq:reheat_S}
{\cal L}_{S} &=&  S \Bigg[ \dfrac{|Y_S|^2\, m_{3/2}}{\kappa_1} (\tilde{S_2} \tilde{S}_2^* +\tilde{N} \tilde{N}^*) + ( Y_S Y_\nu^* \tilde{S_1} \tilde{L}^* H_u^*+ \lambda_3 Y_\nu^*\sqrt{\dfrac{\mu}{M_P}}  \tilde{S_2} \tilde{L}^* H_u^*+ h.c.) \nonumber\\
&&+   2 \kappa_1\,\kappa_2 \, \mu \tilde{S}_1 \tilde{S}_2
\Bigg] 
+ \left[ S \left( Y_S\, \bar{N}^c \, S_1^c  + \lambda_3\sqrt{\dfrac{\mu}{M_P}} \, \bar{N}^c \, S_2^c \right) + \, h.c. \right]
\eea 
\begin{figure}[h!]
	\centering
\begin{overpic}[width=0.6\textwidth]{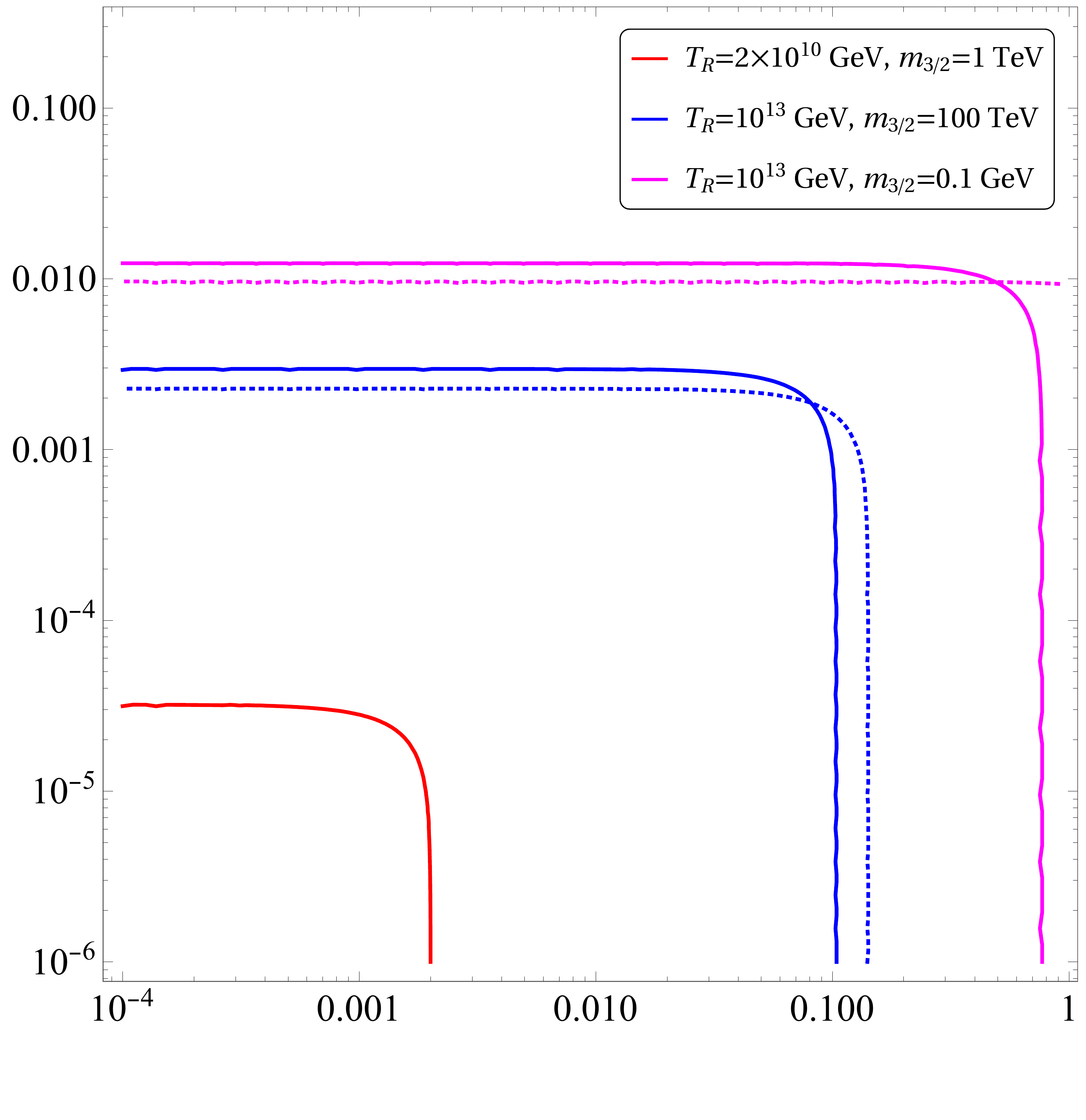}
 \put (55,2) {$\displaystyle \lambda_3 $}
 \put (-5,53) {$\displaystyle Y_S $}
\end{overpic}
	\caption{The allowed regions in the plane $(\lambda_3,Y_S)$,  corresponding to constraints on reheating temperature.
	\label{fig:reheat1}}
\end{figure}	
In case of TeV scale gravitinos, $T_R\lesssim 2\times 10^{10}$ GeV hence $\Gamma \lesssim  10^{3}$ GeV.
Studying the branching ratios of different decay channels with respect to the total decay width, one finds that for  large $Y_S$, the dominant decay channel is $S  \to \, \bar{N}^c \, S_1^c$, hence  a constraint on $Y_S < 10^{-4}$. In this case, the channel 
$S  \to \, \bar{N}^c \, S_2^c$, $\phi_i \to \tilde{S_i} \tilde{S}_j^*$ become the dominant. The latter introduces the constraints $\lambda_3 < 10^{-3}$ and $\kappa_2 < 0.01$. As advocated above, these constraints can be relaxed if gravitinos were light $ m_{3/2} \lesssim {\cal O} $(0.1 GeV) or very heavy $ m_{3/2} \gsim {\cal O} $(100 TeV).
 
In Fig. \ref{fig:reheat1}, we show the allowed region under the curves, for the couplings $Y_S$ and $\lambda_3$, where we have fixed $\mu$ from the scalar perturbations observations and $Y_\nu \sim {\cal O}(1)$.  The red curve corresponds to $T_R\sim 2\times 10^{10}$ GeV with $\kappa_1 \sim 10^{-5}$, $\kappa_2 \sim 10^{-3}$ and  $\tau_0 \sim 10^{-3} \,M_P$. The solid blue(purple) curves correspond to the relaxation $T_R\sim 10^{13}$ GeV, when gravitinos are too heavy or too light, with the respective values $\kappa_1 \sim 10^{-2}( 10^{-3})$, $\kappa_2 \sim 10^{-2}( 10^{-1})$ and  $\tau_0 \sim 1 (10) \,M_P$. For the dashed we have changed $\tau_0 \sim 0.1 (0.01) \,M_P$ respectively.
%
\subsection{Neutrino masses}
\label{subsec:neutrino}
The Lagrangian of the neutrino masses can be derived from the superpotential (\ref{eq:suppot_nu}), after electroweak symmetry breaking, as follows
\bea\label{eq:lag_nu}
{\cal L}_{\nu}= m_D \, \bar{\nu}_L \, N^c+ M_{R_1}\, \bar{N}^c \, S_1^c +   M_{R_2}\, \bar{N}^c \, S_2^c + \mu_S \, \bar{S_1}^c \,S_2^c + h.c.,
\eea
 \bea\label{eq:numassparam}
{\text{with}}\, \hspace{0.5cm} m_D &=&  Y_\nu \, v \sin(\beta) \,,\, 
\hspace{1cm} M_{R_1}  =    \dfrac{Y_S \, m_{3/2}}{\kappa_1} \,, 
\hspace{1cm} M_{R_2}  =    \dfrac{\lambda_3 \, m_{3/2} \, M}{\kappa_1 \, M_P} \,,\nonumber \\
 \mu_S &=& \dfrac{\lambda_2 \,}{M_P}  \,  \langle S \rangle ^2 \, + \, \dfrac{\kappa_2 \, }{M_P} \left[\langle \phi_1 \phi_2 \rangle - M^2\right]  \nonumber \\
 %
 &=& \dfrac{  m_{3/2}^2}{\kappa_1^2 \, M_P} \left[ \lambda_2 \, - \, \kappa_2   \, + \,  \dfrac{ \kappa_2 \, m_{3/2}^2}{2\kappa_1^2 \, M^2} \,  \right] \,,
 \eea 
where $v$ is the electroweak vev and $\tan\beta=\dfrac{\langle H_u \rangle}{\langle H_d \rangle}$. 
The last term in $\mu_S$ is suppressed by the super-heavy scale $M^2$ and dominant effects come from the first two terms.
 The neutrino mass matrix, in the basis $(\nu_L, N, S_1, S_2)$ (for one flavor), is given by 
\begin{figure}[b!]
	\centering
\begin{overpic}[width=0.49\textwidth]{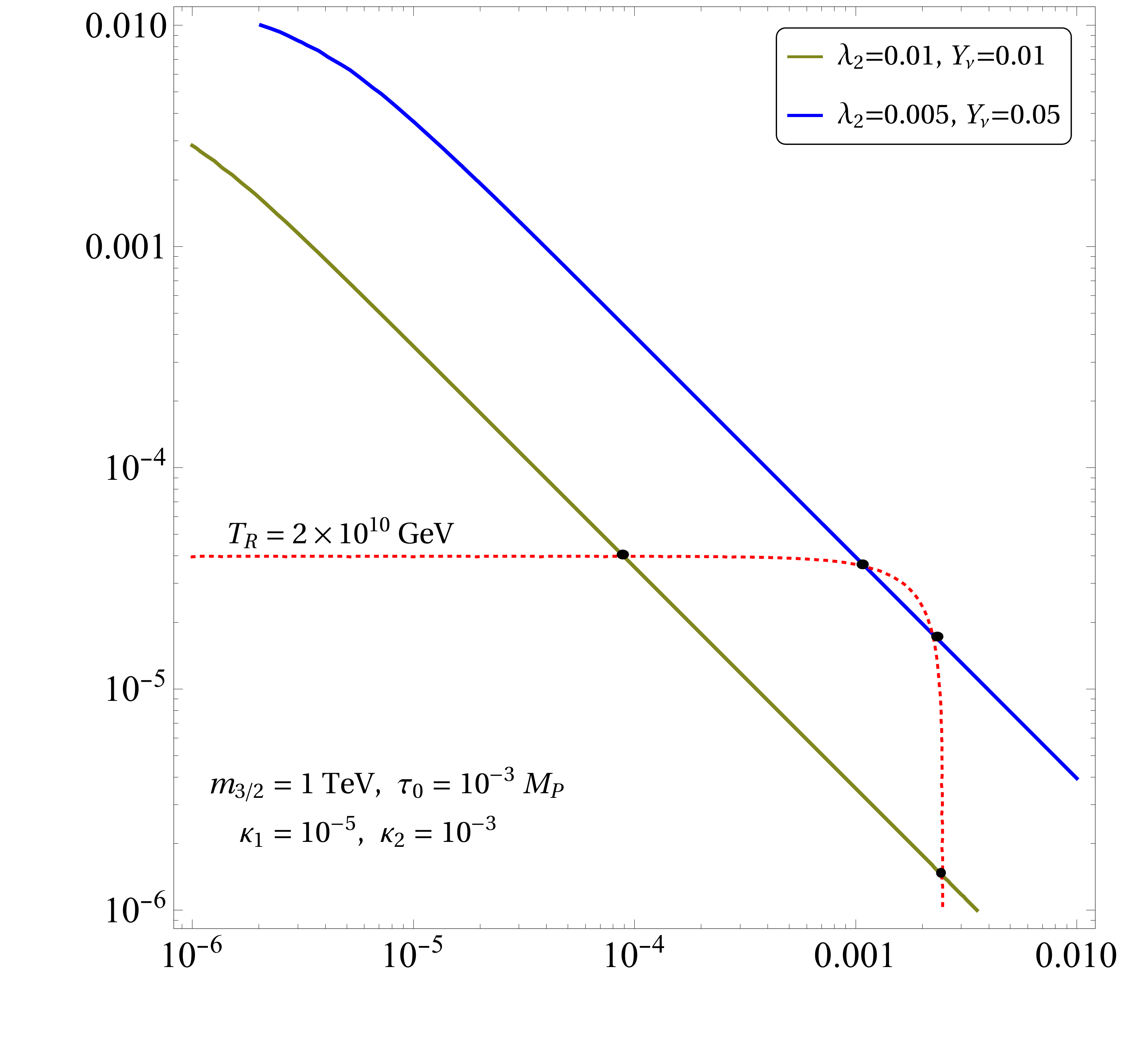}
 \put (55,2) {$\displaystyle \lambda_3 $}
 \put (1,50) {$\displaystyle Y_S $}
\end{overpic}
\begin{overpic}[width=0.49\textwidth]{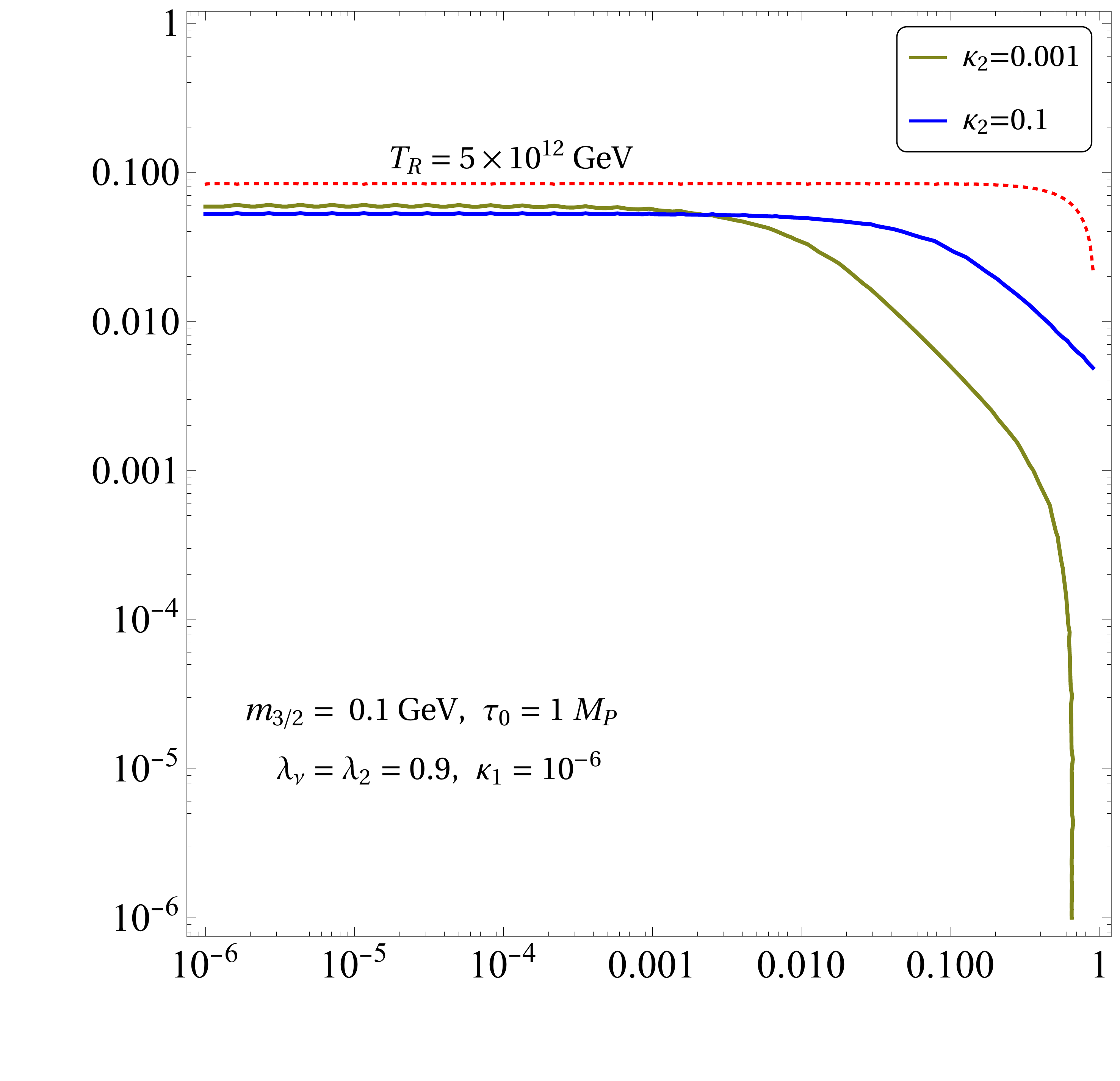}
 \put (55,2) {$\displaystyle \lambda_3 $}
 \put (1,55) {$\displaystyle Y_S $}
\end{overpic}
\begin{overpic}[width=0.49\textwidth]{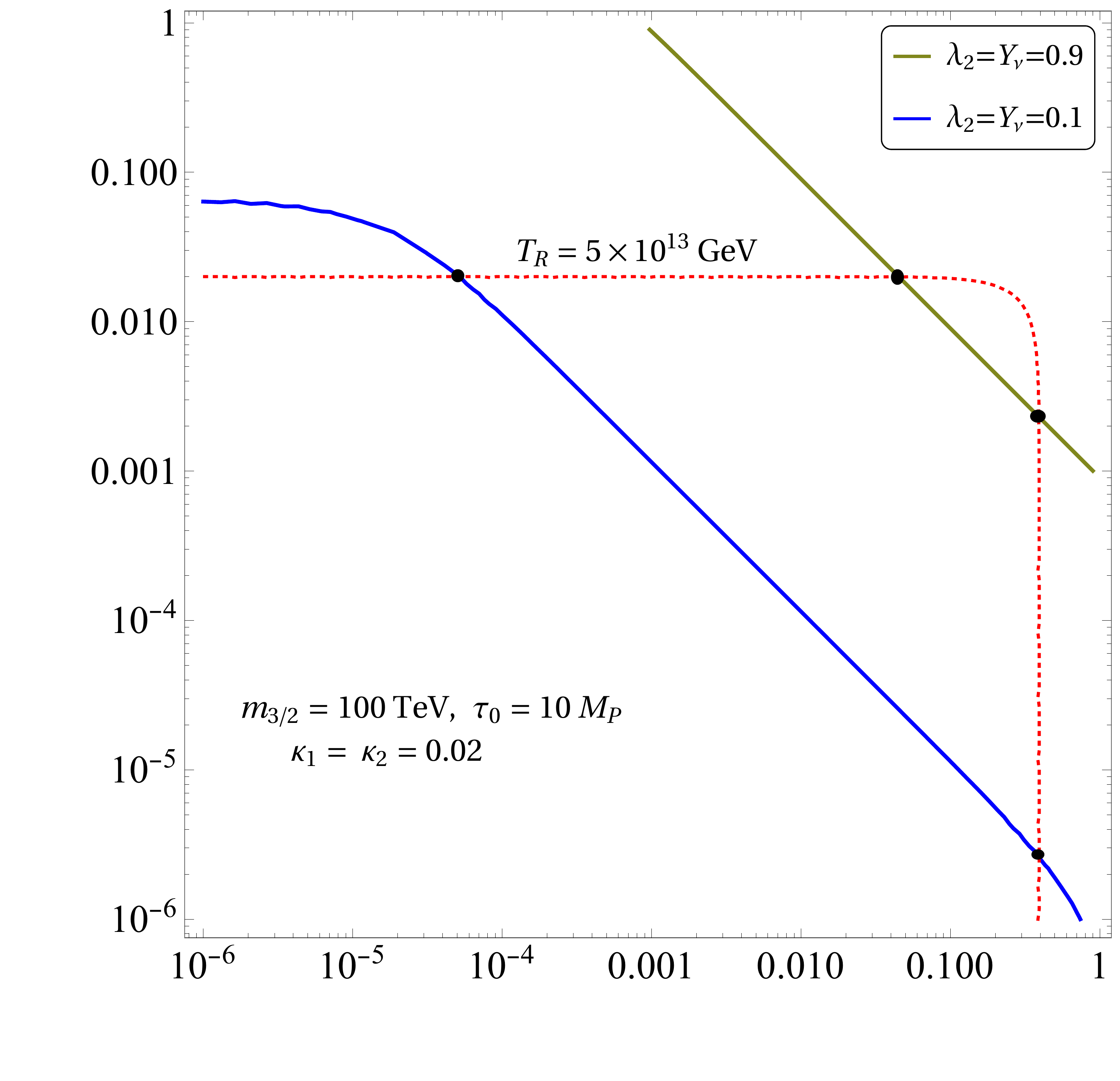}
 \put (55,2) {$\displaystyle \lambda_3 $}
 \put (1,55) {$\displaystyle Y_S $}
\end{overpic}
	\caption{The regions of parameter space in the plane $  (\lambda_3 , Y_S) $ allowed by neutrino masses and reheating temperature constraints. The solid lines represent the lightest neutrino eigenstate $| m_{\nu_l} | \sim {\cal O}(0.1)$ eV, while the dotted curve represent the constraints from reheating.
	\label{fig:mnu}}
\end{figure}
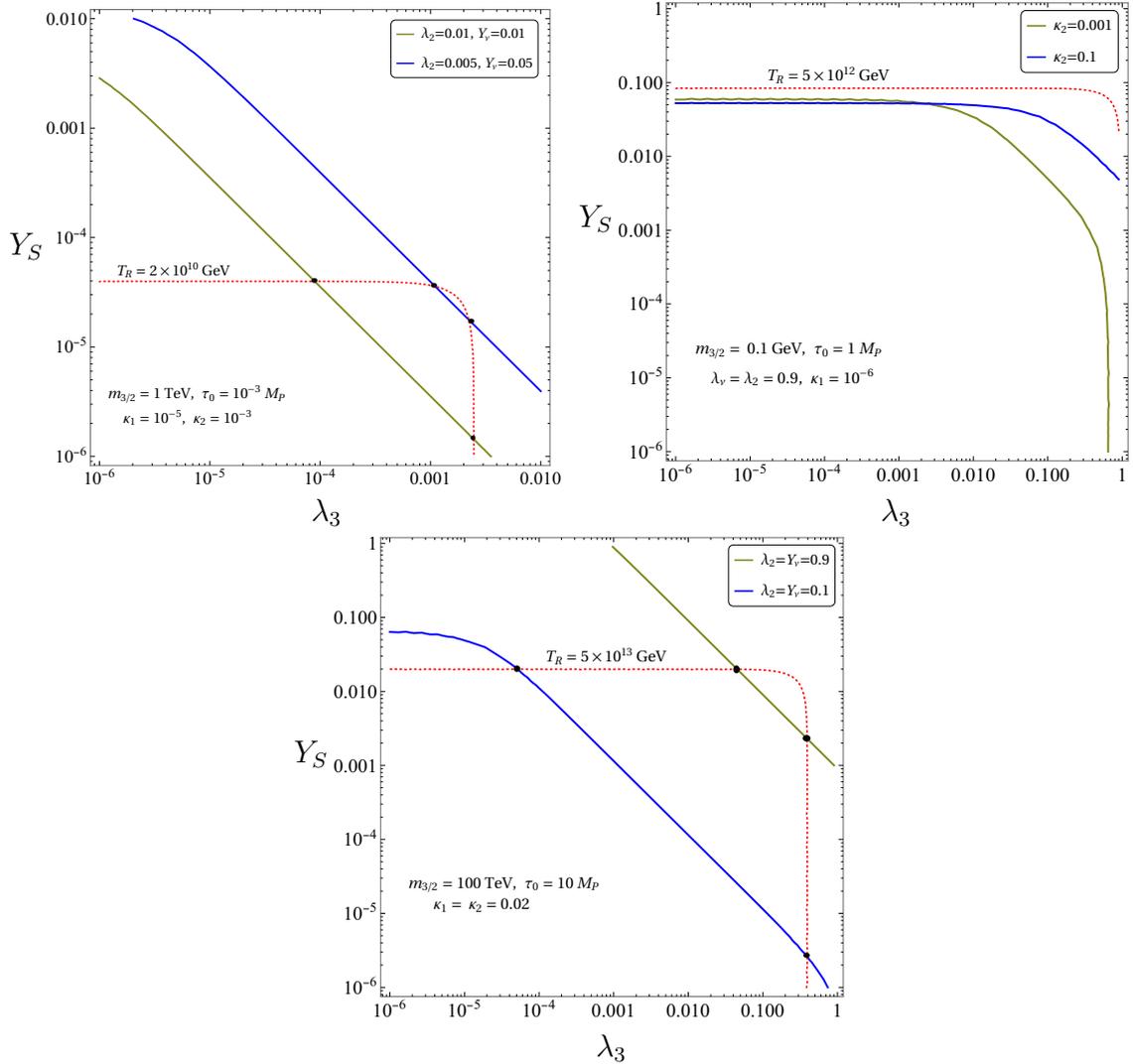	
\bea\label{eq:mass_nu}
{M}_{\nu}= \left(
\begin{array}{cccc}
 0 & {m_D} & 0 & 0  \\
 {m_D} & 0 & {M_{R_1}} & M_{R_2} \\
 0 & {M_{R_1}}  & 0 & \mu_S  \\
 0  & M_{R_2} & \mu_S   & 0 \\
\end{array}
\right) .
\eea
For the mass hierarchy $\mu_S\ll m_D \ll M_{R_i}$, the eigenvalues consist of a light  mass corresponding to the active neutrino, a relatively light one corresponding a sterile neutrino and two heavy states as follows 
\bea
m_{\nu_{l}}&\simeq & \frac{m_D^2 \,\mu _S}{2 M_{R_1} M_{R_2}}\,, \nonumber \\
m_{\nu_{1}}&\simeq & \frac{\mu _S \left(-\frac{m_D^2}{2 M_{R_2}}-2 M_{R_2}\right)}{M_{R_1}} \, \sim \,  \mu _S \,, \nonumber \\
m_{\nu_{2,3}}&\simeq &  \pm \left(M_{R_1}+\frac{m_D^2+M_{R_2}^2}{2 M_{R_1}}\right)+\frac{M_{R_2} \mu _S}{M_{R_1}} \sim \pm \left(M_{R_1}+M_{R_2}\right).
\eea
In Table \ref{tab:nuparam}, we list three points in the parameter space that give neutrino mass of order ${\cal O}(0.1)$ eV, where upper bound on the sum over the neutrino
masses is given by $\sum_\gamma m_{\nu_\gamma} < 0.15$ eV, at 95\% CL \cite{Giusarma:2016phn,RoyChoudhury:2018gay}. We have taken different value of $m_{3/2}$ corresponding to different regimes in the reheating scenario and fixed $ \sin(\beta) \sim 1$. The value of  $ M $ can range between $ 10^{15}-10^{17}$ GeV. 
\begin{table}[h]
 \centering
 \begin{tabular}{c c c c c | c c | c c }
 \hline \hline
  $ \,\,\,\, m_{3/2} \,\,\,\, $ & $\,\,\,\, \lambda_2 \,\,\,\, $ & $\,\,\,\, \lambda_3 \,\,\,\, $  & $ \,\,\,\,  \kappa_1 \,\,\,\,  $ & $ \,\,\,\,  \kappa_2 \,\,\,\, $  & $\,\,\,\, Y_\nu \,\,\,\, $ & $\,\,\,\, Y_S \,\,\,\,  $ & $\,\,\,\, | m_{\nu_1} | \,\,\,\,  $ & $\,\,\,\,  |m_{\nu_{2,3}}|  \,\,\,\,  $ 
  \\
   \hline 
   $10^3$ & 0.008 & $0.002 $ & $10^{-5} $ & $10^{-3} $&  0.05 & $3\times 10^{-5} $ & $3\times 10^{-5} $  &  $ 4.4 \times 10^{3} $ 
   \\
   $0.1$ & 0.9 & 0.4 & $10^{-7} $   & 0.1 & 0.9 &  0.01 & $2.1 \times 10^{-7} $ &  $1.06 \times 10^{4} $
   \\
   $10^{5}$ & 0.1 & 0.01 & 0.01 & $0.01 $   & 0.5 & 0.0032  & $6.6 \times 10^{-7} $ &  $3.21 \times 10^{4} $
   \\
   \hline  \hline
  \end{tabular}
 \caption{{\footnotesize  Three benchmark points in the parameter space that give the active neutrino masses $| m_{\nu_l} | \sim {\cal O}(0.1)$ eV and satisfy the constraints from inflation observables and reheating. All mass scales are given in GeV.}}
 \label{tab:nuparam}
\end{table}
Fig. \ref{fig:mnu} depicts the allowed values of the parameters in the plane $  (\lambda_3 , Y_S) $, while fixing the other parameters, where the solid curves correspond to active neutrino mass value $| m_{\nu_l} | \sim {\cal O}(0.1)$ eV. We have fixed $\sin(\beta) \sim 1$, and $Y_\nu \sim {\cal O}(1)$ in the case where $m_{3/2}= 0.1 ,10^5$ GeV. We have imposed the constraints from inflation observables and reheating temperature according to the cases discussed in (\ref{subsec:reheat}). Therefore the segments between the black dots give the allowed values in the parameter space $  (\lambda_3 , Y_S) $ for fixed values of the other parameters.
%
%
\section{Conclusions}
\label{sec:conclusions}
We have constructed an inflation model in which the non-singlet right-handed sneutrinos $S_1,S_2$ contain the inflaton component, in no-scale supergravity. We have modified the tribrid inflation model to implement a scenario in which the inflation sector is linked to a TeV inverse seesaw mechanism. In this regard, we have explained the smallness of the $\mu_S$ mass parameter on a basis of the symmetry defined on the model and SUSY breaking effects. We have discussed different regimes in the parameter space that lead to successful inflation compatible with Planck observation. We found that the reheating scenario imposes severe constraints on the parameter space when gravitino mass is TeV scale. These constraints can be relaxed for very heavy or ultra-light gravitinos in order to avoid the BBN constraints. We have derived the neutrino masses taking into account the latter constraints. 
The model may have interesting consequences on the low energy phenomenology.

\acknowledgments
The author would like to thank Stefan Antusch and Waleed Abdallah for encouragement and stimulating discussions. 
The work of A. M. is  supported in part by the Science, Technology and Innovation Funding Authority (STDF) 
under grant No. (33495) YRG.



\begin{thebibliography}{99}

\bibitem{Akrami:2018odb}
Y.~Akrami \textit{et al.} [Planck],
``Planck 2018 results. X. Constraints on inflation,''
Astron. Astrophys. \textbf{641}, A10 (2020)
[arXiv:1807.06211 [astro-ph.CO]].

\bibitem{Super-Kamiokande:1998kpq}
Y.~Fukuda \textit{et al.} [Super-Kamiokande],
``Evidence for oscillation of atmospheric neutrinos,''
Phys. Rev. Lett. \textbf{81}, 1562-1567 (1998)
[arXiv:hep-ex/9807003 [hep-ex]].

\bibitem{LSND:2001aii}
A.~Aguilar-Arevalo \textit{et al.} [LSND],
``Evidence for neutrino oscillations from the observation of $\bar{\nu}_e$ appearance in a $\bar{\nu}_\mu$
 beam,''
Phys. Rev. D \textbf{64}, 112007 (2001)
[arXiv:hep-ex/0104049 [hep-ex]].

\bibitem{K2K:2002icj}
M.~H.~Ahn \textit{et al.} [K2K],
``Indications of neutrino oscillation in a 250 km long baseline experiment,''
Phys. Rev. Lett. \textbf{90}, 041801 (2003)
[arXiv:hep-ex/0212007 [hep-ex]].

\bibitem{T2K:2011ypd}
K.~Abe \textit{et al.} [T2K],
``Indication of Electron Neutrino Appearance from an Accelerator-produced Off-axis Muon Neutrino Beam,''
Phys. Rev. Lett. \textbf{107}, 041801 (2011)
[arXiv:1106.2822 [hep-ex]].

\bibitem{DayaBay:2013yxg}
F.~P.~An \textit{et al.} [Daya Bay],
``Spectral measurement of electron antineutrino oscillation amplitude and frequency at Daya Bay,''
Phys. Rev. Lett. \textbf{112}, 061801 (2014)
[arXiv:1310.6732 [hep-ex]].


\bibitem{Starobinsky:1980te} 
  A.~A.~Starobinsky,
  ``A New Type of Isotropic Cosmological Models Without Singularity,''
  Phys.\ Lett.\  {\bf 91B}, 99 (1980);
  V.~F.~Mukhanov and G.~V.~Chibisov,
  ``Quantum Fluctuations and a Nonsingular Universe,''
  JETP Lett.\  {\bf 33}, 532 (1981)
  [Pisma Zh.\ Eksp.\ Teor.\ Fiz.\  {\bf 33}, 549 (1981)];
  A.~A.~Starobinsky,
  ``The Perturbation Spectrum Evolving from a Nonsingular Initially De-Sitter Cosmology and the Microwave Background Anisotropy,''
  Sov.\ Astron.\ Lett.\  {\bf 9}, 302 (1983).
%




\bibitem{Cremmer:1983bf}
E.~Cremmer, S.~Ferrara, C.~Kounnas and D.~V.~Nanopoulos,
``Naturally Vanishing Cosmological Constant in N=1 Supergravity,''
Phys. Lett. B \textbf{133} (1983), 61.

\bibitem{Ellis:1984bm}
J.~R.~Ellis, C.~Kounnas and D.~V.~Nanopoulos,
``No Scale Supersymmetric Guts,''
Nucl. Phys. B \textbf{247} (1984), 373-395.

\bibitem{Ellis:2013xoa} 
  J.~Ellis, D.~V.~Nanopoulos and K.~A.~Olive,
  ``No-Scale Supergravity Realization of the Starobinsky Model of Inflation,''
  Phys.\ Rev.\ Lett.\  {\bf 111}, 111301 (2013)
  Erratum: [Phys.\ Rev.\ Lett.\  {\bf 111}, no. 12, 129902 (2013)]
  [arXiv:1305.1247 [hep-th]].
%

%
\bibitem{Ellis:2013nxa}
J.~Ellis, D.~V.~Nanopoulos and K.~A.~Olive,
``Starobinsky-like Inflationary Models as Avatars of No-Scale Supergravity,''
JCAP \textbf{10} (2013), 009 
[arXiv:1307.3537 [hep-th]].

\bibitem{Ellis:2017jcp}
J.~Ellis, M.~A.~G.~Garcia, N.~Nagata, D.~V.~Nanopoulos and K.~A.~Olive,
``Starobinsky-like Inflation, Supercosmology and Neutrino Masses in No-Scale Flipped SU(5),''
JCAP \textbf{07} (2017), 006
[arXiv:1704.07331 [hep-ph]].

\bibitem{Khalil:2018iip}
S.~Khalil, A.~Moursy, A.~K.~Saha and A.~Sil,
``$U(1)_R$ inspired inflation model in no-scale supergravity,''
Phys. Rev. D \textbf{99} (2019) no.9, 095022
[arXiv:1810.06408 [hep-ph]].

\bibitem{Moursy:2020sit}
A.~Moursy,
``No-scale hybrid inflation with R-symmetry breaking,''
JHEP \textbf{02}, 230 (2021)
[arXiv:2009.14149 [hep-ph]].

\bibitem{Civiletti:2013cra}
M.~Civiletti, M.~Ur Rehman, E.~Sabo, Q.~Shafi and J.~Wickman,
``R-symmetry breaking in supersymmetric hybrid inflation,''
Phys. Rev. D \textbf{88} (2013) no.10, 103514
[arXiv:1303.3602 [hep-ph]].


\bibitem{Gonzalo:2016gey}
T.~E.~Gonzalo, L.~Heurtier and A.~Moursy,
``Sneutrino driven GUT Inflation in Supergravity,''
JHEP \textbf{06} (2017), 109
[arXiv:1609.09396 [hep-th]].


\bibitem{Dvali:1994ms}
  G.~R.~Dvali, Q.~Shafi and R.~K.~Schaefer,
  ``Large scale structure and supersymmetric inflation without fine tuning,''
  Phys.\ Rev.\ Lett.\  {\bf 73}, 1886 (1994).
  [hep-ph/9406319].
%

\bibitem{Murayama:1992ua}
H.~Murayama, H.~Suzuki, T.~Yanagida and J.~Yokoyama,
``Chaotic inflation and baryogenesis by right-handed sneutrinos,''
Phys. Rev. Lett. \textbf{70}, 1912-1915 (1993).


\bibitem{Antusch:2004hd}
S.~Antusch, M.~Bastero-Gil, S.~F.~King and Q.~Shafi,
``Sneutrino hybrid inflation in supergravity,''
Phys. Rev. D \textbf{71}, 083519 (2005)
[arXiv:hep-ph/0411298 [hep-ph]].

\bibitem{Antusch:2008pn}
S.~Antusch, M.~Bastero-Gil, K.~Dutta, S.~F.~King and P.~M.~Kostka,
``Solving the eta-Problem in Hybrid Inflation with Heisenberg Symmetry and Stabilized Modulus,''
JCAP \textbf{01}, 040 (2009)
[arXiv:0808.2425 [hep-ph]].

\bibitem{Antusch:2009vg}
S.~Antusch, K.~Dutta and P.~M.~Kostka,
``Tribrid Inflation in Supergravity,''
AIP Conf. Proc. \textbf{1200}, no.1, 1007-1010 (2010)
[arXiv:0908.1694 [hep-ph]].

\bibitem{Antusch:2010va}
S.~Antusch, M.~Bastero-Gil, J.~P.~Baumann, K.~Dutta, S.~F.~King and P.~M.~Kostka,
``Gauge Non-Singlet Inflation in SUSY GUTs,''
JHEP \textbf{08}, 100 (2010)
[arXiv:1003.3233 [hep-ph]].

\bibitem{Antusch:2012jc}
S.~Antusch and D.~Nolde,
``K\"ahler-driven Tribrid Inflation,''
JCAP \textbf{11}, 005 (2012)
[arXiv:1207.6111 [hep-ph]].

\bibitem{Antusch:2013eca}
S.~Antusch and F.~Cefal\`a,
``SUGRA New Inflation with Heisenberg Symmetry,''
JCAP \textbf{10}, 055 (2013)
[arXiv:1306.6825 [hep-ph]].

\bibitem{Minkowski:1977sc}
P.~Minkowski,
``$\mu \to e\gamma$ at a Rate of One Out of $10^{9}$ Muon Decays?,''
Phys. Lett. B \textbf{67}, 421-428 (1977)

\bibitem{Mohapatra:1979ia}
R.~N.~Mohapatra and G.~Senjanovic,
``Neutrino Mass and Spontaneous Parity Nonconservation,''
Phys. Rev. Lett. \textbf{44}, 912 (1980)

\bibitem{Yanagida:1979as}
T.~Yanagida,
``Horizontal gauge symmetry and masses of neutrinos,''
Conf. Proc. C \textbf{7902131}, 95-99 (1979)
KEK-79-18-95.

\bibitem{Schechter:1980gr}
J.~Schechter and J.~W.~F.~Valle,
``Neutrino Masses in SU(2) x U(1) Theories,''
Phys. Rev. D \textbf{22}, 2227 (1980)

\bibitem{Mohapatra:1986aw}
R.~N.~Mohapatra,
``Mechanism for Understanding Small Neutrino Mass in Superstring Theories,''
Phys. Rev. Lett. \textbf{56}, 561-563 (1986)

\bibitem{Mohapatra:1986bd}
R.~N.~Mohapatra and J.~W.~F.~Valle,
``Neutrino Mass and Baryon Number Nonconservation in Superstring Models,''
Phys. Rev. D \textbf{34}, 1642 (1986)

\bibitem{Gonzalez-Garcia:1988okv}
M.~C.~Gonzalez-Garcia and J.~W.~F.~Valle,
``Fast Decaying Neutrinos and Observable Flavor Violation in a New Class of Majoron Models,''
Phys. Lett. B \textbf{216}, 360-366 (1989)

\bibitem{Kachru:2003aw} 
  S.~Kachru, R.~Kallosh, A.~D.~Linde and S.~P.~Trivedi,
  ``De Sitter vacua in string theory,''
  Phys.\ Rev.\ D {\bf 68}, 046005 (2003),
  [hep-th/0301240].
%
\bibitem{Balasubramanian:2005zx} 
  V.~Balasubramanian, P.~Berglund, J.~P.~Conlon and F.~Quevedo,
  ``Systematics of moduli stabilisation in Calabi-Yau flux compactifications,''
  JHEP {\bf 0503}, 007 (2005)
  [hep-th/0502058]; 
  V.~Balasubramanian, P.~Berglund, J.~P.~Conlon and F.~Quevedo,
  ``Systematics of moduli stabilisation in Calabi-Yau flux compactifications,''
  JHEP {\bf 0503}, 007 (2005)
  [hep-th/0502058].
%

\bibitem{Coleman:1973jx}
S.~R.~Coleman and E.~J.~Weinberg,
``Radiative Corrections as the Origin of Spontaneous Symmetry Breaking,''
Phys. Rev. D \textbf{7}, 1888-1910 (1973)
%

\bibitem{Dvali:1995fb}
G.~R.~Dvali,
``Inflation induced SUSY breaking and flat vacuum directions,''
Phys. Lett. B \textbf{355}, 78-84 (1995)
[arXiv:hep-ph/9503375 [hep-ph]].
%


\bibitem{Dvali:1997uq}
G.~R.~Dvali, G.~Lazarides and Q.~Shafi,
``Mu problem and hybrid inflation in supersymmetric $SU(2)_L \times SU(2)_R \times U(1)_{(B-L)}$,''
Phys. Lett. B \textbf{424} (1998), 259-264
[arXiv:hep-ph/9710314 [hep-ph]].

\bibitem{Barbieri:1982eh}
R.~Barbieri, S.~Ferrara and C.~A.~Savoy,
``Gauge Models with Spontaneously Broken Local Supersymmetry,''
Phys. Lett. B \textbf{119} (1982), 343;
A.~H.~Chamseddine, R.~L.~Arnowitt and P.~Nath,
``Locally Supersymmetric Grand Unification,''
Phys. Rev. Lett. \textbf{49} (1982), 970;
%
H.~P.~Nilles, M.~Srednicki and D.~Wyler,
``Weak Interaction Breakdown Induced by Supergravity,''
Phys. Lett. B \textbf{120} (1983), 346;
L.~J.~Hall, J.~D.~Lykken and S.~Weinberg,
``Supergravity as the Messenger of Supersymmetry Breaking,''
Phys. Rev. D \textbf{27} (1983), 2359-2378;
%
Q.~Shafi, A.~Sil and S.~P.~Ng,
``Hybrid inflation, dark energy and dark matter,''
Phys. Lett. B \textbf{620}, 105-110 (2005)
[arXiv:hep-ph/0502254 [hep-ph]].
%
\bibitem{Buchmuller:2000zm}
W.~Buchmuller, L.~Covi and D.~Delepine,
``Inflation and supersymmetry breaking,''
Phys. Lett. B \textbf{491}, 183-189 (2000)
[arXiv:hep-ph/0006168 [hep-ph]].
W.~Buchm\"uller, V.~Domcke, K.~Kamada and K.~Schmitz,
``Hybrid Inflation in the Complex Plane,''
JCAP \textbf{07}, 054 (2014)
doi:10.1088/1475-7516/2014/07/054
[arXiv:1404.1832 [hep-ph]].

\bibitem{Ellis:1982yb}
J.~R.~Ellis, A.~D.~Linde and D.~V.~Nanopoulos,
``Inflation Can Save the Gravitino,''
Phys. Lett. B \textbf{118}, 59-64 (1982)

\bibitem{Khlopov:1984pf}
M.~Y.~Khlopov and A.~D.~Linde,
``Is It Easy to Save the Gravitino?,''
Phys. Lett. B \textbf{138}, 265-268 (1984)

\bibitem{Moroi:1995fs}
T.~Moroi,
``Effects of the gravitino on the inflationary universe,''
[arXiv:9503210 [hep-ph]].


\bibitem{Ellis:1984eq}
J.~R.~Ellis, J.~E.~Kim and D.~V.~Nanopoulos,
``Cosmological Gravitino Regeneration and Decay,''
Phys. Lett. B \textbf{145} (1984), 181-186
%

\bibitem{Ellis:1984er}
J.~R.~Ellis, D.~V.~Nanopoulos and S.~Sarkar,
``The Cosmology of Decaying Gravitinos,''
Nucl. Phys. B \textbf{259} (1985), 175-188

\bibitem{Moroi:1993mb}
T.~Moroi, H.~Murayama and M.~Yamaguchi,
``Cosmological constraints on the light stable gravitino,''
Phys. Lett. B \textbf{303} (1993), 289-294

\bibitem{Kawasaki:2004yh}
M.~Kawasaki, K.~Kohri and T.~Moroi,
``Hadronic decay of late - decaying particles and Big-Bang Nucleosynthesis,''
Phys. Lett. B \textbf{625} (2005), 7-12
[arXiv:astro-ph/0402490 [astro-ph]].

\bibitem{Antusch:2010mv}
S.~Antusch, J.~P.~Baumann, V.~F.~Domcke and P.~M.~Kostka,
``Sneutrino Hybrid Inflation and Nonthermal Leptogenesis,''
JCAP \textbf{10} (2010), 006
[arXiv:1007.0708 [hep-ph]].

\bibitem{Kawasaki:2008qe}
M.~Kawasaki, K.~Kohri, T.~Moroi and A.~Yotsuyanagi,
``Big-Bang Nucleosynthesis and Gravitino,''
Phys. Rev. D \textbf{78}, 065011 (2008)
[arXiv:0804.3745 [hep-ph]].

\bibitem{Ferrantelli:2010as}
A.~Ferrantelli,
``Gravitino phenomenology and cosmological implications of supergravity,''
[arXiv:1002.2835 [hep-ph]].

\bibitem{Ferrara:2010in}
S.~Ferrara, R.~Kallosh, A.~Linde, A.~Marrani and A.~Van Proeyen,
``Superconformal Symmetry, NMSSM, and Inflation,''
Phys. Rev. D \textbf{83}, 025008 (2011)
[arXiv:1008.2942 [hep-th]].

\bibitem{Giudice:2003jh}
G.~F.~Giudice, A.~Notari, M.~Raidal, A.~Riotto and A.~Strumia,
``Towards a complete theory of thermal leptogenesis in the SM and MSSM,''
Nucl. Phys. B \textbf{685}, 89-149 (2004)
[arXiv:hep-ph/0310123 [hep-ph]].

\bibitem{Pradler:2006hh}
J.~Pradler and F.~D.~Steffen,
``Constraints on the Reheating Temperature in Gravitino Dark Matter Scenarios,''
Phys. Lett. B \textbf{648}, 224-235 (2007)
[arXiv:hep-ph/0612291 [hep-ph]].

\bibitem{Ellis:2015jpg}
J.~Ellis, M.~A.~G.~Garcia, D.~V.~Nanopoulos, K.~A.~Olive and M.~Peloso,
``Post-Inflationary Gravitino Production Revisited,''
JCAP \textbf{03}, 008 (2016)
[arXiv:1512.05701 [astro-ph.CO]].

\bibitem{Ellis:2016spb}
J.~Ellis, H.~J.~He and Z.~Z.~Xianyu,
``Higgs Inflation, Reheating and Gravitino Production in No-Scale Supersymmetric GUTs,''
JCAP \textbf{08}, 068 (2016)
[arXiv:1606.02202 [hep-ph]].

\bibitem{Dudas:2017rpa}
E.~Dudas, Y.~Mambrini and K.~Olive,
``Case for an EeV Gravitino,''
Phys. Rev. Lett. \textbf{119}, no.5, 051801 (2017)
[arXiv:1704.03008 [hep-ph]].

\bibitem{Ellis:2020lnc}
J.~Ellis, M.~A.~G.~Garcia, N.~Nagata, N.~D.~V., K.~A.~Olive and S.~Verner,
``Building models of inflation in no-scale supergravity,''
Int. J. Mod. Phys. D \textbf{29}, no.16, 2030011 (2020)
[arXiv:2009.01709 [hep-ph]].

\bibitem{Giusarma:2016phn}
E.~Giusarma, M.~Gerbino, O.~Mena, S.~Vagnozzi, S.~Ho and K.~Freese,
``Improvement of cosmological neutrino mass bounds,''
Phys. Rev. D \textbf{94}, no.8, 083522 (2016)
[arXiv:1605.04320 [astro-ph.CO]].

\bibitem{RoyChoudhury:2018gay}
S.~Roy Choudhury and S.~Choubey,
``Updated Bounds on Sum of Neutrino Masses in Various Cosmological Scenarios,''
JCAP \textbf{09}, 017 (2018)
[arXiv:1806.10832 [astro-ph.CO]].



\end{thebibliography}
\end{document}